%% file: IFAC_Attack_Review_Final_Submission.tex
\documentclass[conference]{ifacconf}

\usepackage{cite}
\usepackage{natbib}    % required for bibliography
\usepackage{algorithmic}
\usepackage{textcomp}
\usepackage{xcolor}
\def\BibTeX{{\rm B\kern-.05em{\sc i\kern-.025em b}\kern-.08em
		T\kern-.1667em\lower.7ex\hbox{E}\kern-.125emX}}

\usepackage{graphics} % for pdf, bitmapped graphics files
\usepackage{epsfig} % for postscript graphics files
\usepackage{mathptmx} % assumes new font selection scheme installed
\usepackage{times} % assumes new font selection scheme installed
\usepackage{amsmath} % assumes amsmath package installed
\usepackage{amssymb} % assumes amsmath package installed
\usepackage{graphicx} 
\usepackage{cite}
\usepackage{bbold}
\usepackage{float}
\usepackage{circuitikz}
\usepackage{tikz}
\usepackage{caption}
\usepackage{dsfont}
\usepackage{bm}

\input{macros}

\newcommand{\be}{\begin{equation}}
\newcommand{\ee}{\end{equation}}
\newcommand{\beqn}{\begin{eqnarray}}
\newcommand{\eeqn}{\end{eqnarray}}
\newcommand{\ba}{\begin{array}}
	\newcommand{\ea}{\end{array}}

\newcommand{\R}{\mathds{R}}

\newcommand{\Vs}{\nu }
\newcommand{\ep}{\hfill$\blacksquare$}

\newcommand{\nquad}{\mkern-18mu}
\newcommand{\nqquad}{\mkern-36mu}

\newcommand{\tr}{\textnormal{tr}}
\usetikzlibrary{shapes,arrows}
\usetikzlibrary{positioning}

\tikzstyle{block} = [draw, fill=RR_black!0, rectangle, minimum height=2em, minimum width=6em]
\tikzstyle{rect1} = [rectangle, minimum height=3.1em, minimum width=3.1em]
\tikzstyle{rect2} = [rectangle, minimum height=3.1em, minimum width=1em]
\tikzstyle{rect3} = [fill=RR_black!50,rectangle, minimum height=2em, minimum width=2em]
\tikzstyle{comm1} = [draw,dashed,fill=RR_black!10, rectangle, minimum height=4em, minimum width=16em]
\tikzstyle{comm} = [draw, dashed, fill=RR_black!10, rectangle, minimum height=5em, minimum width=16em]
\tikzstyle{rect} = [draw, fill=RR_black!0, rectangle, minimum height=2em, minimum width=2em]
\tikzstyle{sum} = [draw, fill=RR_black!70, circle]
\tikzstyle{input} = [coordinate]
\tikzstyle{output} = [coordinate]
\tikzstyle{pinstyle} = [pin edge={to-,thin,RR_black}]

\definecolor{PU_orange}{RGB}{245,128,37}
\definecolor{PU_orange_light}{RGB}{245,178,78}
\definecolor{RR_blue}{RGB}{16,6,159}
\definecolor{RR_blue_light}{RGB}{0,109,255}
\definecolor{RR_black}{RGB}{30,54,67}
\definecolor{RR_pink}{RGB}{250,70,146}
\definecolor{RR_paragraph}{RGB}{78,93,101}
\definecolor{RR_white}{RGB}{255,255,255}

\begin{document}
	
	%===============================================================================
	
	\begin{frontmatter}
		
		\title{Denial of Service Attacks on Control Systems with Packet Loss}

		
		\thanks{This work is supported by  Rolls-Royce, ESPRC, and The Control, Monitoring and Systems Engineering UTC at The University of Sheffield.}
		
		\author[First]{William Casbolt}
		\author[First,Second]{I\~{n}aki Esnaola}
		\author[First]{Bryn Jones}
		
		\address[First]{Department of Automatic Control and Systems Engineering, University of Sheffield, Sheffield S1 3JD, UK, (e-mail: \{wgcasbolt1, esnaola, b.l.jones\}@sheffield.ac.uk ).}
		\address[Second]{Department of  Electrical Engineering, Princeton University, Princeton, NJ 08544, USA.}
%		\address[Third]{Bryn Jones is with the Faculty of Department Control and Systems Engineering, University of Sheffield.
%			 	{\tt\small b.l.jones@sheffield.ac.uk}}
		
		\begin{abstract}        % Abstract of not more than 250 words.

	% Update IID section to explicitly mention detection.
	
	 The performance of control systems with packet loss as a result of an attack over the actuation communication channel is analysed. The operator is assumed to monitor the state of the channel by measuring the average number of packet losses and an attack detection criteria is established based on the statistic. The performance of the attacker is measured in terms of the increase of the linear quadratic cost function of the operator subject to a given detection constraint. Within that setting, the optimal denial of service (DoS) attack strategy is formulated for UDP-like and TCP-like communication protocols. {For both communication protocols,} DoS attack constructions that are independent and identically distributed (IID) are compared to those that are non-stationary. The main contributions of this paper are (i) explicit characterisation of the expected cost increase of the optimal attack constructions and the associated packet loss parameter for the IID case, (ii) proof, by example, that non-stationary random attacks outperform IID attacks in the presence of detection constraints.

		\end{abstract}
		
		
	\end{frontmatter}

	%===============================================================================

\section{Introduction}
The introduction of advanced sensing and communication capabilities to control systems gives rise to vulnerabilities that can be exploited with a malicious intent by an attacker (\cite{SCADA}). While the security challenges that control systems face are multifaceted and of diverse nature, the simplicity of implementation of Denial of Service (DoS) attacks in field zones and control zones makes them particularly suitable to exploit software and hardware faults. In this paper we study DoS attacks over control systems that experience packet losses over an actuator communication channel.
%
 %Figure. \ref{fig1} and Fig. \ref{fig2} . 
%
To account for packet loss, we consider the two protocols proposed in \cite{1}. The first one in which the packet loss is not monitored, termed UDP-like for its similarity to the communication protocol. The second protocol is termed as TCP-like and monitors the packet loss realisation by sending a packet receipt acknowledgement message back to the receiver. Both protocols and the systems they constitute are depicted in Fig. \ref{fig1} and Fig. \ref{fig2}. 
In the literature these are termed {\it like} due to previous transmissions being monitored and not retransmitted~(\cite{1}),~(\cite{4}),~(\cite{7}),~(\cite{8}). In \cite{paper1} the authors consider a deterministic DoS attack strategy on a system with power constraints whereas we consider a random DoS attack construction that operates within a no-detection region using an MPC formulation. We first study attack sequences that are constructed as an independent and identically distributed (IID) process. The rationale for this attack construction stems from the simplicity of the attack implementation and the robust attack performance for a wide range of system parameters. We then propose a non-stationary random attack construction and show that for some systems their dynamics can be exploited by the attacker to improve upon the IID construction. 

%how a system using these protocols performs when subjected to independent and identically distributed (IID) denial of service (DoS) attacks. This is done using the MPC framework. In order to derive the optimal attack, the problem is split into two perspectives. The first perspective is the system operator, who uses the optimal control strategy derived in~(\cite{MPC}), and the other is the attacker who maximises the operators cost function with respect to the mean of the loss variable, $\Mm$. The attacker attempts to maximise the cost function whilst remaining undetected by the system operator. Initially, the attacker analyses the system in an unconstrained setting (no detection criteria) and the optimal attack strategy is derived for both protocols. In doing so, it is shown that an optimal attack strategy exists that is IID in nature for both protocols. The attacker then restricts the attack to guarantee a smaller cost increase without violating the detection criteria. From there, constructions of attacks that are not Identically distributed but still independent are investigated. The optimal attack for the non-stationary attacks are formed as a quadratic programming (QP) problem for each protocol.
%
\section{Plant Model}
%
%The operator of the control system implements the optimal control scheme laid out in~(\cite{MPC}). It is made clear that the optimal control law for both protocols includes weighting terms that depend on the probability of packet losses. A natural question therefore, is how sensitive is the closed loop system to a DoS attack? Specifically the attacker aims to increase the cost of the system for the operator through control of the probability of a packet loss in the actuation channel. 
%%

%%
%The system for the operator is modelled as a Gauss Markov process given by,
We consider the plant model given by
\vspace{-3mm}
\beqn\label{eq1}
X_{k+1}&=&{{{\bf A}}} X_k + {{\bf B}}{{\bf V}_k}U_k +W_k, 
\eeqn 
where ${{{\bf A}}}\in \R^{n\times n}$ is the dynamics matrix, $X_k \in \R^n$ describes the state of the plant at time step $k\in\mathds{N}$,
%and $N$ denotes the prediction horizon
${{\bf B}} \in \R^{n \times m}$ is the 
%%
%controls
control 
matrix, ${U}_k\in \R^m$ is the vector of control inputs at the $k$-th time step, $W_k \in \R^n$ is the process noise modelled by a Gaussian distributed vector of random variables with mean ${\bf 0} \in \R^n$ and covariance matrix 
%%
% ${ \Sigma_{XX}} \in S^n_{+}$
%%
${ \Sigma_{W}} \in S^n_{++}$ where, $S^n_{++}$ is the set of $n\times n$ symmetric positive definite matrices,
and ${\bf V}_k \in S^{m}_{+}$ is the packet loss variable where the $i$-th diagonal entry is an IID Bernoulli random variable with mean $\mu_i$. We assume that the current state of the plant is determined by the vector of Gaussian distributed random variables $X_k$ with mean ${\overline{X}\in\mathds{R}^n}$ and covariance matrix ${ \Sigma_{X}} \in S^n_{++}$.

%%
%$\left(\PP\left[{\bf V}_{i;k}=1\right]=\mu_{i}\right)$ where it is clear the semicolon separates the position index from the time. 
%%
%%
%%
%$X_k$ is normally distributed such that $X_k \sim \mathcal{N}\left(\overline{X} , {\bf P}\right)$, $\overline{X}\in\R^{n}$ is the mean of $X_k$ and ${\bf P}\in \Sm^{n}_{+}$ is the covariance matrix, ${U}_k\in \R^m$ is the control input vector at time instance $k$, ${{{\bf A}}}\in \R^{n\times n}$ is the dynamics matrix, ${{\bf B}} \in \R^{n \times m}$ is the controls matrix, $W_k \in \R^n$ is a Gaussian distributed independent random variable with mean ${\bf 0} \in \R^n$ and covariance matrix ${ \Sigma_{W}} \in S^n_{+}$, and ${\bf V}_k \in \Sm^{m}_+$ is assumed to be a diagonal matrix where each entry is an IID Bernoulli random variable with mean $\Mm_i$ $\left(\PP\left[{\bf V}_{i;k}=1\right]=\Mm_i\right)$ where it is clear the semicolon separates the position index from the time. This sequence is IID and therefore the expected value is time-invariant and the index is dropped $\left(\Mm_{i;k}=\Mm_i\ \forall\ k\right)$.  All singular sub-indexes indicate the time instance unless specified otherwise.
%%

\begin{figure}[!t]
	\captionsetup{justification=centering,margin=2cm,width=\linewidth}
	\centering
	\includegraphics[scale=1]{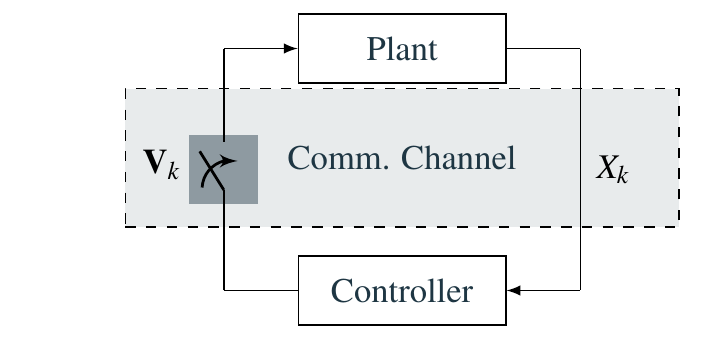}
	\caption{The UDP-like protocol where the realisation of the packet loss ${\bf V}_k$ is not monitored}\label{fig1}
\end{figure}

\begin{figure}[!t]
	\captionsetup{justification=centering,margin=2cm,width=\linewidth}
	\centering
	\includegraphics[scale=1]{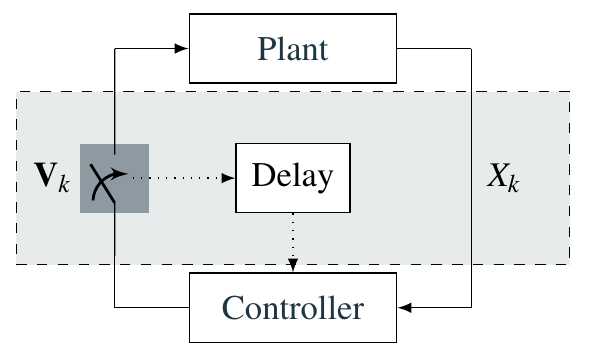}
	\caption[center]{The TCP-like protocol where the realisation of the packet loss ${\bf V}_k$ is transmitted back to the controller.}\label{fig2}
\end{figure}

In this paper, we adopt the MPC formulation used in \cite{MPC} to describe the plant model in (\ref{eq1}) over the prediction horizon $N\in\mathds{N}_+$, resulting in the prediction model given by
\vspace{-3mm}
 \beqn\label{eq2}
\chi_k^{}=\Phi X_k +\Gamma {\Vs}_k{\Upsilon_k} +\Lambda {\Xi_k},
\eeqn 
where $\Phi \in \R^{Nn\times n}$ is the dynamics matrix over the prediction horizon, $\chi_k^{} \in \R^{Nn}$ is the state prediction vector, $\Gamma \in \R^{Nn\times Nm}$  is the propagation matrix for the control law over the prediction horizon, ${\Upsilon_k} \in \R^{mN}$ is the realisation at the $k$-th time step of the control law, $\Lambda \in \R^{Nn\times Nn}$ is the propagation matrix for the process noise, ${\Xi_k} \in \R^{Nn}$ is the process noise over the prediction horizon, and $\Vs_k$ is a diagonal matrix with the Bernoulli random variables describing the packet losses over the prediction horizon along the diagonal. All the terms in (\ref{eq2}) are presented in (\ref{big}).
%%
%where $\chi_k^{},{\Xi_k^{}} \in \R^{Nn}$, $\Phi \in \R^{Nn\times n}$, $\Gamma \in \R^{Nn\times Nm}$, $\Lambda \in \R^{Nn\times Nn}$ $\Vs \in\Sm^{Nm}_+$, and ${\Upsilon_k} \in \R^{mN}$. All terms are described explicitly in (\ref{big}). Note that $\Vs_k$ is an idempotent matrix, therefore ${\Vs^{}}^i=\Vs$ for all $i \in \mathbb{Z}_{++}$, due to the matrix entries being Bernoulli distributed. In each protocol the amount of information available changes the error prediction outcome. 
%
%The information available to the operator for estimation and control is represented by the information sets $\mathcal{F}_k$ and $\mathcal{G}_k$.The information sets are defined formally as:
%%
Due to the lossy communication between the controller and the plant, the operator implements a communication protocol to monitor the state of the packets transmitted to the plant. We adopt the two protocol paradigms proposed by \cite{1}, namely a UDP-like protocol that does not monitor the channel and a TCP-like protocol that acknowledges receipt of the packet from the controller by sending an {\it acknowledgement} message to the controller over an auxiliary channel. The difference between both protocol paradigms is depicted in Fig. \ref{fig1} and Fig. \ref{fig2}, which show the UDP-like and the TCP-like protocols, respectively. The information set available to the controller is determined by the choice of the protocol. We define the information sets as
\be
\Ic_k=
\begin{cases}
\Fc_k=\left\lbrace X^k, \Vm^{k-1}\right\rbrace,&\mbox{TCP-like,}\\
\Gc_k=\left\lbrace X^k\right\rbrace,&\mbox{UDP-like},\\
\end{cases}
\ee
%\beqn\label{eq3}
%\mathcal{I}_k &=& \left\{ \ba{cc} \mathcal{F}_k=\left\{X_k, \Vs^{k-1} \right\},&\qquad\mbox{TCP-like} \\ \mathcal{G}_k=\left\{X_k\right\},\qquad \ \ \ &\qquad\mbox{UDP-like}\hfill \ea\right.
%\eeqn 
%
%%
where $\Vm^{k-1}=\left\lbrace \Vm_0, \Vm_1, \ldots, \Vm_{k-1}\right\rbrace$, $X^{k}=\left\lbrace X_0, X_1, \ldots, X_{k}\right\rbrace$ and all sets are monotonically increasing, i.e. there is a filtration such that $\mathcal{I}_{k} \subseteq \mathcal{I}_{k+1}$. It is shown in \cite{MPC} that for both protocols the optimal control law is determined by the mean of the packet loss variable $\bar{\Vs}\eqdef \mathds{E}[\Vs_k]$. Following in the steps of \cite{1}, the performance of the controller is characterised by a linear quadratic Gaussian (LQG) cost function. The description of the cost function in the MPC framework proposed in \cite{MPC} is the cost function
%%
%where $\Vs^{k-1} = \left(\ba{cc} \Vs^{k-2} &{\bf 0} \\ {\bf 0} & {\bf V}_{k-1} \ea\right) $ and all sets are monotonically increasing, $\mathcal{I}_{k} \subseteq \mathcal{I}_{k+1}$. As seen in~(\cite{MPC}), the optimal control law depends on the estimation of the mean of the loss variable, $\Mm$, for both protocols. Naturally, as an attacker of this system it is beneficial to characterise how altering the statistics of the variable ${\bf V}_k$ alters the overall cost of the operator. In order to analyse this, the operator uses the same linear quadratic Gaussian (LQG) cost function as in~(\cite{MPC}): 
%%
%
%%
\vspace{-5mm}
 \beqn\label{eq4}
\!\!\! J^*\!\!\left(\Ic_k\right)\triangleq \min_{\Upsilon_k}\left\{\!\mathds{E}\left[X_k^{^{\sf T}} \Qm X_k + \chi_k^{^ {\sf T}} \Omega \chi_k^{} + {\Upsilon_k}^{\sf T} \Vs^{^{\sf T}}_k {\Psi} \Vs_k {\Upsilon_k}\big|\mathcal{I}_k\right]\!\right\}\!\!,
\eeqn 
%%
% \beqn\label{eq4}
%\mathds{E}\left[J^*(X_k,{\Upsilon_k},\Vs_k) \big|\mathcal{I}_k\right]&\triangleq& \nonumber \\
%&& \nqquad\nqquad\nqquad\min_{{\Upsilon_k}}\left\{\mathds{E}\left[X_k^{^{\sf T}} \Qm X_k + \chi_k^{^ {\sf T}} \Omega \chi_k^{} + {\Upsilon_k}^{\sf T} \Vs^{^{\sf T}}_k {\Psi} \Vs_k {\Upsilon_k}\big|\mathcal{I}_k\right]\right\},
%\eeqn 
%%
where $\Omega\in S_{++}^{Nn}$ is the state penalty diagonal matrix, ${\Psi}\in S_{++}^{Nm}$ is the input penalty diagonal matrix, and the diagonal matrix $\Qm \in S_{++}^{n}$. The optimal control law for~(\ref{eq4}) is obtained in~\cite{MPC} for each protocol and shown to be
%%
%where the function is weighted with the diagonal matrices denoted by $\Omega\in\Sm_{++}^{Nn\times Nn}$ (state penalty matrix), ${\Psi}\in \Sm_{++}^{Nm\times Nm}$ (input penalty matrix). The optimal minimising input of~(\ref{eq4}) is the same as in~(\cite{MPC})
%%
%\beqn
%\label{eq:Upsilon}
%\Upsilon_{k|\mathcal{I}_k}^*&=& \left\{\ba{cc}  \Upsilon_{k|\mathcal{F}_k}^*& \qquad\mbox{TCP-like} \\
% \Upsilon_{k|\mathcal{G}_k}^* & \qquad\mbox{UDP-like} \ea \right., \\
%  \Upsilon_{k|\mathcal{F}_k}^*&=&-\left({\Psi} + {\Delta}^{\Gamma}\overline{\Vs}\right)^{-1}\Fm X_k, \nonumber \\
% \Upsilon_{k|\mathcal{G}_k}^*&=&-\left({\Psi} + {\Delta}^{\Gamma}\overline{\Vs} +\left({\bf I} \odot {\Delta}^{\Gamma}\right)\left(1 -\overline{\Vs}\right) \right)^{-1}\Fm X_k,\nonumber
%\eeqn
\vspace{-3mm}
\beqn
\label{eq:Upsilon}
\Upsilon_{k|\!\mathcal{I}_k}^*\!\!\!\!=\!\! \left\{\ba{cc} \!\!\!\!\!\!\!\!\!\!\!\!\!\!\!\!\!\!\!\!\!\!\!\!\!\!\!\!\!\!\!\!\!\!\!\!\!\!\!\!\!\!\!\!\!\!\! \Upsilon_{k|\mathcal{F}_k}^*\!\!=\!\!-\left(\!{\Psi} + {\Delta}^{\Gamma}\overline{\Vs}\right)^{-1}\Fm X_k, \qquad &\!\! \mbox{TCP-like,} \\
\!\!\!\! \Upsilon_{k|\mathcal{G}_k}^*\!\!=\!\!-\left(\!{\Psi} + {\Delta}^{\Gamma}\overline{\Vs} +\left(\!{\bf I} \odot {\Delta}^{\Gamma}\right)\!\!\left(\!1 -\overline{\Vs}\right) \right)^{-1}\!\!\Fm X_k, &\!\! \mbox{UDP-like,} \ea \right.\!\!\!\!\!\!\!\!\!\!\! \nonumber 
\\
\eeqn
where $\odot$ is the element-wise Hadamard product. 
\subsection{Attack Model}
%%
%
%==============================================================================================================
%
%
%%
The performance of the controller is determined by the mean of the packet losses in the actuation channel. In view of this, we study the security risk posed by an attacker that governs the statistics of the packet losses on the actuation channel. In practice, this can be achieved by the attacker via DoS attacks over the communication channel. We are not concerned with the particular implementation of the DoS attacks, instead we study the packet loss attack strategy that aims to disrupt the operation of the controller. In particular, we consider the case in which the attacker constructs the attack sequence by designing a random distribution. The rationale for this stems from the fact that the operator expects the packet losses to be IID, and therefore, the attacker mimics the {\it nominal} operation of the channel. That being the case, the optimal attack construction is characterised by the probability of packet loss in the actuation channel, described by the diagonal matrix ${\bf V}^\alpha_k \in S^{m}_{++}$ where the $i$-th diagonal entry is an IID Bernoulli random variable with mean $\mu^\alpha_i$. 

\begin{figure*}[!t]
	\fontsize{10pt}{2pt}
	\begin{align}\label{big}\!\!\!
	\underset{\chi_k^{}}{\underbrace{\left(\ba{c}X_{k+1} \\ X_{k+2} \\ \vdots \\ X_{k+N}
			\ea\right)}}\!\!\!
	= \!\!\!
	\underset{\Phi}{\underbrace{\left(\ba{c}
			{{{\bf A}}} \\ {{\bf A}}^2 \\ \vdots \\ {{{\bf A}}}^N 
			\ea\right)}}X_k\!
	+ \!\!\!
	\underset{\Gamma}{\underbrace{\left(\ba{cccc}
			{\bf B} &\bf 0 &\dots & \bf 0\\
			{{\bf A}}{{\bf B}} & {{\bf B}}& \ddots& \vdots\\
			\vdots& \ddots& \ddots& \bf 0 \\
			{{\bf A}}^{N-1}{\bf B} & \dots & {{\bf A}} {\bf B} & {\bf B}
			\ea\right)}}\
	\underset{\Vs_k}{\underbrace{\left(\ba{cccc}
			{\bf V}_k &\bf 0 &\dots & \bf 0 \\
			\bf 0 & {\bf V}_{k+1} & \ddots & \vdots \\
			\vdots & \ddots & \ddots & \bf 0 \\
			\bf 0 &\dots & \bf 0 & {\bf V}_{k+N-1}
			\ea \right)}}\ 
	\underset{{\Upsilon_k}}{\underbrace{\left(\ba{c}
			U^n_k\\
			U^n_{k+1}\\
			\vdots\\
			U^n_{k+N-1}
			\ea\right)}}\!\!
	+\!\!
	\underset{\Lambda}{\underbrace{\left(\ba{cccc}
			{\bf I} &\bf 0 &\dots & \bf 0\\
			{{\bf A}} & {\bf I} & \ddots& \vdots\\
			\vdots& \ddots& \ddots& \bf 0 \\
			{{\bf A}}^{N-1} & \dots & {{\bf A}} & {\bf I}
			\ea\right)}}\
	\underset{\Xi_k}{\underbrace{\left(\ba{c}
			W_k \\
			W_{k+1}\\
			\vdots \\
			W_{k+N-1}
			\ea\right)}}\!\!\!\!\!\!\!\!
	\end{align}
	\hrulefill
	\vspace*{4pt}
\end{figure*}

To achieve this, the attacker has knowledge of the information set given by
\vspace{-0.3cm}
\beqn\label{eq5}
\mathcal{A}_k &=& \left\{{{\bf A}},{{\bf B}},\Sigma_{W},\bar{\nu}, \Omega, {\Psi}, \Ic_k \right\}.
\eeqn 
It is shown later that knowledge of the state of the plant is not necessary to construct the optimal attack and is only required to compute the cost induced by the attack for a particular realisation of the state variables. 
%%
%Due to the control law depending on properties of $\Vm_k$, it is beneficial, as an attacker of the system, to characterise how controlling the statistics of ${\bf V}_k$ alters the overall cost for the operator.
The controller operates under the assumption that the packet losses over the actuation channel are IID with a mean defined by $\Mm\eqdef\mathds{E}[\Vm_k]$ for $k\in\mathds{N}$, with $\Mm\in S^m_{++}$. By changing the statistics of the actuation channel, the attacker induces a different distribution over the sequence of packet losses. To distinguish the induced case from the nominal case the induced sequence of packet losses is defined as the diagonal matrix ${\bf V}^\alpha_k$ as above. Similarly, the channel is characterised by $\Mm^\alpha\eqdef\mathds{E}[\Vm^\alpha_k]$ for $k\in\mathds{N}$, with $\Mm^\alpha\in S^m_{++}$.
Note that the mean does not depend on the time step $k$, and therefore, the sequence of random variables describing the packet loss in the $i$-th position is IID.
%%
%, i.e. $\{(\Vm_k^\alpha)_{i,i}\}_{k=1}^N$ and for $i=1,2, \ldots, m$.
%%
The sequence of packet losses over the prediction horizon is described by the diagonal matrix $\Vs_k^\alpha$ with the Bernoulli sequences along the diagonal and $\bar{\Vs}^\alpha\triangleq \mathds{E}[\Vs^\alpha_k]$ for $k=1,2,\ldots,N$.
%
 %The information available to the attacker is assumed to contain all design parameters. 
%%
%We define this information set as
%%
%When the system is under attack the random variable controlling the packet losses is not be equivalent to ${\bf V}_k$ and therefore to avoid confusion this random variable is defined as ${\bf V}^\alpha_k \in \Sm^{m}_+$ and is a diagonal matrix where each entry is an IID Bernoulli random variable with mean $\Mm^\alpha_i$ ($\PP\left[{\bf V}^{\alpha}_{i;k}=1\right]=\Mm^{\alpha}_i$, meaning the mean of the $i^th$ actuator is time-invariant). The attacker is assumed to have access to all design parameters of the system, and is represented by the information set $\mathcal{A}_k$ and is defined formally as:
%%
%\beqn\label{eq5}
%\mathcal{A}_k &=& \left\{{{\bf A}},{{\bf B}},\Sigma_{W},\bar{\nu}, \mathcal{I}_k \right\}.
%\eeqn 
%%%
%from this information set the attacker can construct all prediction matrices, (\ref{big}), and the optimal control law that the operator is implementing, $\Upsilon_{k|\mathcal{I}_k}^*$. An IID attack construction is the first considered. This attack reveals how the control system is affected by a change in the received proportion of packet losses. The attacker has no control over the $k^{th}$ input supplied by the operator, $\Upsilon_{k|\mathcal{I}_k}^*$. 

The objective of the attacker, in contrast to the objective of the controller, is to maximise the cost function~(\ref{eq4}). The cost function of the attacker is
%Taking these factors into account the cost function of the attacker is:
%%
%from this information set the attacker has access to all the prediction matrices, (\ref{big}), and the optimal control law that the operator is implementing, $\Upsilon_{k|\mathcal{I}_k}^*$. The first attack construction considered is an IID attack. Doing this reveals how the control system is affected by a change in the expected proportion of packet losses. Note that the attacker has no control over the $k^{th}$ input supplied by the operator, $\Upsilon_{k|\mathcal{I}_k}^*$. This is equivalent to assuming the attacker has access to the controller output and is `reading' the optimal output. The objective of the attacker unlike traditional control, is to maximise the cost function~(\ref{eq4}) as opposed to minimise. Taking these factors into account the cost function of the attacker is:
%%
%%
\vspace{-3mm}
\beqn
 J_A\left(\Ac_k\right)&\triangleq&
\min_{\Upsilon_{k|\mathcal{I}_k}^{*}}\left\{\mathds{E}\left[ X_k^{\sf T} \Qm X_k+\chi_k^{\sf T} \Omega \chi_k+\Upsilon_k^{\sf T}{\Vs}^{\alpha{\sf T}}_k {\Psi} {\Vs}^{\alpha}_k \Upsilon_k\big|\mathcal{A}_k\right]\right\}\nonumber,
\eeqn
%\begin{subequations}\label{eq6}
%\beqn
%J^*_A\left(\Ac_k\right)\triangleq&& \nonumber \\
%&&\nqquad\nqquad\nqquad \ \ \max_{\bar{\Vs}^\alpha}\left\{\min_{\Upsilon_{\mathcal{I}_k}^{*}}\left\{\!\mathds{E}\left[ X_k^{\sf T} \Qm X_k + \chi_k^{\sf T} \Omega \chi_k + \Upsilon_k^{\sf T} {\Vs}^{\alpha{\sf T}}_k {\Psi} {\Vs}^{\alpha}_k \Upsilon_k\big|\mathcal{A}_k\right]\right\}\right\},\ \ \\
%	&& \nqquad \mbox{s.t.}\ \bar{{\Vs}}^{\alpha}_{i,k} \in \Cc\left(\Mm,\Lm\right)\qquad \forall \ \ i={1,\dots,m}, \ \ k=1,\dots,N.
%\eeqn
%\end{subequations}
%%
%\begin{subequations}\label{eq6}
	%\beqn
	%\nqquad\mathds{E}\left[J^*_A(X_k,\Upsilon_k,\Vs^\alpha_k)\big|\mathcal{A}_k\right]\triangleq&& \nonumber \\
	%&& \nqquad\nqquad \nqquad\nqquad\nqquad \max_{\bar{\Vs}^\alpha}\left\{\min_{\Upsilon_{\mathcal{I}_k}^{*}}\left\{\mathds{E}\left[ X_k^{\sf T} \Qm X_k + \chi_k^{\sf T} \Omega %\chi_k + \Upsilon_k^{\sf T} {\Vs}^{\alpha{\sf T}}_k {\Psi} {\Vs}^{\alpha}_k \Upsilon_k\big|\mathcal{A}_k\right]\right\}\right\},\\
%	&& \nqquad\nqquad \nqquad\nqquad\nquad \mbox{s.t.}\ \bar{{\Vs}}^{\alpha}_{i,k} \in \Cc\left(\Mm,\Lm\right)\qquad \forall \ \ i={1,\dots,m}, \ \ k=1,\dots,N.
%	\eeqn
%\end{subequations}
%%
%
%
%=======================================================================================================================
%%
%where $\Cc\left(\Mm,\Lm\right)=\left[0,1\right]$. 
and the optimal attack construction is defined as 
\vspace{-3mm}
\beqn\label{eq6}
J^*_A\left(\Ac_k\right)\triangleq\max_{\bar{\Vs}^\alpha}\left\{J_A\left(\Ac_k\right)\right\}.
\eeqn
Note that, the cost function of the operator in (\ref{eq4}) is nested inside (\ref{eq6}), i.e. the attacker chooses the worst case packet loss mean under the assumption that the controller operates optimally. The state estimation performed by the attacker accounts for the true statistics of the actuation channel to produce the state prediction
%%
%where $\Cc\left(\Mm,\Lm\right)=\left[0,1\right]$. The cost function in~(\cite{MPC}), (\ref{eq4}), is nested inside the maximisation of (\ref{eq6}). The expectation in~(\ref{eq6}) is taken with the knowledge the attacker has access to. The state estimation performed by the attacker is,
%%
\vspace{-5mm}
\beqn\label{eq8}
\hat{\chi}_{k}^{\alpha} &\eqdef& \mathds{E}\left[\chi_k\big|\mathcal{A}_k\right] = \Phi X_k +\Gamma \bar{\Vs}^\alpha \Upsilon_k ,
\eeqn
and the state error prediction of the attack construction for the two protocols is given by
\vspace{-3mm}
\begin{subequations}
	\beqn
	\Em_{k|\mathcal{F}_k}^\alpha &\eqdef& \chi_k^{} - \mathds{E}\left[\chi_k^{}\big|\mathcal{F}_k, \Vs^{\alpha}_k\right] 
	%			 &=& \Phi X_k +\Gamma {\Vs}{\Upsilon_k} + \Lambda {\Xi_k^{}} - \Phi X_k -\Gamma {\Vs}{\Upsilon_k}, \nonumber\\
	= \Lambda {\Xi_k}, \label{eq9a}\\
	\Em_{k|\mathcal{G}_k}^\alpha &\eqdef& \chi_k^{} - \mathds{E}\left[\chi_k^{}\big|\mathcal{G}_k\right]
	%		   &=& \Phi X_k +\Gamma {\Vs}{\Upsilon_k} + \Lambda {\Xi_k^{}} - \Phi X_k -\Gamma \bar{\Vs}{\Upsilon_k}, \nonumber\\
	= \Gamma\left(\Vs^\alpha_k - \bar{\Vs}^\alpha\right){\Upsilon_k} + \Lambda {\Xi_k^{}},
	\label{eq9b}
	\eeqn 	
\end{subequations}
for the TCP-like protocol and the UDP-like protocol, respectively. Note that the TCP-like prediction includes the knowledge of $\Vs^{\alpha}_k$, this is the same knowledge the operator has when calculating this expectation. Specifically, the acknowledgement link removes the effect of the actuation from the estimation. Even though at time step $k$ the realisation of $\Vm_{k+1}$ is not known, it is known at time step $k+1$, and therefore, the operator knows that any chosen actuation law does not affect the estimation at future time steps as this contribution can be removed from the estimation. This is not possible if the acknowledgement channel is not a perfect channel.
The proof of the error trajectories is analogous to the proof in~\cite{MPC} and is omitted. 
We describe the attack induced cost by rewriting (\ref{eq6}) in terms of the state prediction and the state prediction error as in~\cite{MPC} which yields
%Expanding (\ref{eq6}) into the state error plus the estimated state as in~(\cite{MPC}) yields
%%
\vspace{-3mm}
\beqn
J^*_A\left(\Ac_k\right)&=&\max_{\bar{\Vs}^\alpha}\left\{\min_{\Upsilon_{k|\mathcal{I}_k}^{*}}\left\{\mathds{E}\left[ \vphantom{\Big |}X_k^{\sf T} \Qm X_k +  \Upsilon_k^{\sf T}{\Vs}^{\alpha{\sf T}}_k {\Psi}{\Vs}^{\alpha}_k\Upsilon_k  \right.\right. \right.\nonumber \\%Not strictly necessary
&& \qquad\quad\left.\left. \left.+ \left(\hat{\chi}_{k}^{\alpha}+\Em_k^{\alpha} \right)^{\sf T} \Omega \left(\hat{\chi}_{k}^{\alpha}+\Em_k^{\alpha} \right) {\Big|}\mathcal{A}_k\right]\right\}\right\},\nonumber \\
%
%&&\nqquad\nqquad =X_k^{\sf T} \Qm X_k + \max_{\bar{\Vs}^\alpha}\left\{\min_{\Upsilon_{\mathcal{I}_k}^{*}}\left\{\hat{\chi}_{k}^{\alpha{\sf T}} \Omega \hat{\chi}_{k}^{\alpha}+\Upsilon_k^{\sf T} \bar{\Vs}^\alpha {\Psi} \Upsilon_k +\mathds{E}\left[ \Em_k^{\alpha{\sf T}} \Omega \Em_k^{\alpha} \big|\mathcal{A}_k\right]\right\}\right\},\nonumber \\
%
&& \nquad =X_k^{\sf T}\left( \Qm + {\Delta}^\Phi \right) X_k + \max_{\bar{\Vs}^\alpha}\left\{ \min_{\Upsilon_{k|\mathcal{I}_k}^{*}}\left\{\mathds{E}\left[ \Em_k^{\alpha{\sf T}} \Omega \Em_k^{\alpha} {\Big|}\mathcal{A}_k\right]\right\}\right.\nonumber \\ 
&& \left. \quad +  \Upsilon_{k|\mathcal{I}_k}^{*{\sf T}}\bar{\Vs}^\alpha \left( 2\Fm X_k + \left({\Delta}^\Gamma \bar{\Vs}^\alpha +{\Psi} \right)\Upsilon_{k|\mathcal{I}_k}^*\right) \right\},\label{eq11}
\eeqn
%%
%\beqn
%J^*_A\left(\Ac_k\right)&=&\max_{\bar{\Vs}^\alpha}\left\{\min_{\Upsilon_{\mathcal{I}_k}^{*}}\left\{\mathds{E}\left[ \vphantom{\Bigg |}X_k^{\sf T} \Qm X_k + \left(\hat{\chi}_{k}^{\alpha}+\Em_k^{\alpha} \right)^{\sf T} \Omega \left(\hat{\chi}_{k}^{\alpha}+\Em_k^{\alpha} \right) \right.\right. \right.\nonumber \\
%&& \qquad\qquad\qquad\left.\left. \left.+ \Upsilon_k^{\sf T}{\Vs}^{\alpha{\sf T}}_k {\Psi}{\Vs}^{\alpha}_k\Upsilon_k\big|\mathcal{A}_k\vphantom{\Bigg |}\right]\right\}\right\} %\label{eq10}\\
%
%&&\nqquad\nqquad =X_k^{\sf T} \Qm X_k + \max_{\bar{\Vs}^\alpha}\left\{\min_{\Upsilon_{\mathcal{I}_k}^{*}}\left\{\hat{\chi}_{k}^{\alpha{\sf T}} \Omega %\hat{\chi}_{k}^{\alpha}+\Upsilon_k^{\sf T} \bar{\Vs}^\alpha {\Psi} \Upsilon_k +\mathds{E}\left[ \Em_k^{\alpha{\sf T}} \Omega \Em_k^{\alpha} %\big|\mathcal{A}_k\right]\right\}\right\},\nonumber \\
%
%&&\nqquad\nqquad \nqquad\nqquad\nqquad\nqquad =X_k^{\sf T}\left( \Qm + {\Delta}^\Phi \right) X_k + \max_{\bar{\Vs}^\alpha}\left\{ \Upsilon_{k|\mathcal{I}_k}^{*{\sf T}}\bar{\Vs}^\alpha %\left( 2\Fm X_k + \left({\Delta}^\Gamma \bar{\Vs}^\alpha +{\Psi} \right)\Upsilon_{k|\mathcal{I}_k}^*\right) \right.\nonumber \\
%&& \left. +\min_{\Upsilon_{\mathcal{I}_k}^{*}}\left\{\mathds{E}\left[ \Em_k^{\alpha{\sf T}} \Omega \Em_k^{\alpha} \big|\mathcal{A}_k\right]\right\}\right\},\label{eq11}
%\eeqn
%
where ${\Delta}^{\Phi}=\Phi^{\sf T}\Omega\Phi$, ${\Delta}^{\Gamma}=\Gamma^{\sf T}\Omega\Gamma$, and $\Fm=\Gamma^{\sf T}\Omega\Phi$. 
%In (\ref{eq11}) the optimal $\Upsilon_{k|\mathcal{I}_k}^*$ is substituted.
%
\subsection{Monitoring of Packet Losses and Attack Detection}
%%
%Clarify the nomenclature for Plant, controller, system. What do each know? What are the tasks? 
The optimal control law for both protocols is determined by the mean of packet losses as shown in (\ref{eq:Upsilon}). In view of this, the operator monitors the average number of losses on the actuation channel to check that it agrees with the postulated statistic used to construct the control law. Given that the packet losses form a Bernoulli IID sequence, the distribution is fully characterised by the mean of packet losses $\Mm$. To that end, the system computes the average number of packet losses over each dimension on the channel up to time step $k\in\mathds{N}$ thus producing the estimate for dimension $i$ given by
\be
\hat{\mu}_{i,k}=\frac1k\sum_{j=1}^k (\Vm_j)_{i,i},%\mathds{1}_{\{(\Vm_j)_{i,i}=1\}}
\ee
where $(\Vm_j)_{i,i}$ describes the $i,i$-th element of $\Vm_j$. The resulting estimate of the mean probability of packet loss at time step $k$ is given by
\be
\widehat{\Mm}_k=\textnormal{diag}(\hat{\mu}_{1,k}, \hat{\mu}_{2,k}, \ldots, \hat{\mu}_{m,k}).
\ee
The system uses the estimate to check whether the actuation channel is nominal. In this setting, nominal operation entails that the estimated mean does not deviate significantly from the postulated mean used by the controller to implement the control law. Specifically, the operator determines a safe operation region shaped as a hypercube centered around $\Mm$ and with edge lengths determined by $\Lm\in S^m_+$, where the structure of the lengths is such that $\Lm=\textnormal{diag}(\epsilon_1,\epsilon_2, \ldots, \epsilon_m)$ to account for the different detection thresholds $\{\epsilon_i\}_{i=1}^m$ for each dimension of the actuation channel. The resulting safe operation region is given by
\be
\label{eq:region}
%We might change this to definition over vectors \muv of length m and diag(\muv)-\Mm
\Cc(\Mm, \Lm)=\left\{\widehat{\Mm}_k\in \left[0,1\right]^{m\times m}:-\Lm\preceq\widehat{\Mm}_k-\Mm\preceq \Lm\right\}.\ee
In this setting, an attack is declared at time step $k\in\mathds{N}$ if $\widehat{\Mm}_k\notin\Cc(\Mm, \Lm)$. Otherwise, normal operation of the system continues and the operator monitors the packet losses by updating its estimate $\widehat{\Mm}_k$ at time step $k\in\mathds{N}$. Note that the operator does not incorporate a monitoring performance metric in the cost function; instead, the packet loss monitoring procedure operates concurrent to the system operation but independently of the controller. The detection criteria chosen is optimal, the false alarm rate is determined by $\Lm $, if the operator decides upon a time varying $\Lm_k$ instead of a fixed $\Lm$ then instead define the $\Cc(\Mm, \Lm)$ above as the intersection of all the time varying regions. In view of this, the attack construction is concerned with two performance metrics: the cost increase induced by the attack on the performance of the controller and satisfying that the calculated average of the packet losses induced by the attack conforms to the safe region defined by (\ref{eq:region}). In the following, we discuss optimal attack construction strategies. It should be noted that the attacker has access to the variables $\Lm$ and $\Mm_k$.
\section{IID Attack Construction}
% Mention in conclusion that outcome is similar for all cases.
%

As a result of having access to different information sets, the UDP-like protocol error trajectory given in (\ref{eq9b}) depends on the mean of the control variable for the attacker, while the TCP-like protocol does not depend on the mean of the control variable, as shown in (\ref{eq9a}). For that reason, the derivation is presented separately for the each protocol.
\subsection{UDP-like Protocol}
The optimal attack strategy  for the UDP-like protocol is the solution to the optimisation problem
\vspace{-3mm}
\begin{subequations}\label{eq:UDP-opt}
\beqn
	&&\max_{\bar{\Vs}^\alpha}\; \;J_A\left(\Ac_k\right) \\
	&& \mbox{s.t.}\;\quad \Mm^\alpha \in \Cc\left(\Mm,\Lm\right).
\eeqn
\end{subequations}
%\begin{eqnarray}
%\underset{\av}{\textnormal{minimise}}  \quad  & \| \av\|_{\ell_{1}} \\
%\textnormal {subject to}\quad & \Bm_s\av=\yv
%\end{eqnarray}
Note that the maximisation aims to increase the cost incurred by the controller as a result of the packet losses induced by the attack while the constraint aims to keep the attack within the safe operation region. Additionally, the maximisation in (\ref{eq:UDP-opt}) and the minimisation of the control law in~\cite{MPC} differ in that $\Mm^{\alpha}\neq \Mm$, and therefore, the terms within $\Upsilon_{k|\mathcal{G}_k}^*$ described in (\ref{eq:Upsilon}) do not cancel. In the UDP-like setting the information set, $\Ac_k$, does not have access to the previous realisations of packet losses for estimation, that is, $\Ic_k=\Gc_k$ in (\ref{eq5}).
Using Lemma 1 from~\cite{MPC} and (\ref{eq11}) yields the equivalent cost function given by
\vspace{-3mm}
\beqn
\label{eq:basic}
J^*_A\left(\Ac_k\right)&=& X_k^{\sf T}\!\left( \Qm + {\Delta}^\Phi \right) X_k+\tr\left({\Delta}^{\Lambda} \Sigma_{\Xi}\right) +\max_{\bar{\Vs}^\alpha}\!\left\{\! \Upsilon_{k|\mathcal{G}_k}^{*{\sf T}}\bar{\Vs}^\alpha\!\! \left( 2\Fm X_k \right.\right.\nonumber \\
&& \left.\left.+  \left({\Delta}^\Gamma \bar{\Vs}^\alpha +{\Psi} +\left({\bf I}\odot {\Delta}^{\Gamma}\right)\left({\bf I}- \bar{\Vs}^{\alpha}\right)\right)\Upsilon_{k|\mathcal{G}_k}^*\right)\right\},
%&& \mbox{s.t.}\;\quad \Mm^\alpha \in \Cc\left(\Mm,\Lm\right).
\eeqn
where ${\Delta}^\Lambda=\Lambda^{\sf T}\Omega\Lambda$ and the maximisation is subject to $\Mm^\alpha \in \Cc\left(\Mm,\Lm\right)$. In the following, without loss of generality and for the sake of presentation clarity, it is assumed that all actuators for the system share a single communication channel as in~\cite{1}. This simplifies the attack construction while displaying the same properties of the general attack construction. 
That being the case, $\bar{\Vs}^\alpha$ is a diagonal matrix with equal entries, and therefore, $\bar{\Vs}^\alpha=\alpha{\bf I} $ where 
%% Change identity from {\bf I} to {\bf I}
$\alpha \in \R$ is the control variable of the attacker and $\alpha{\bf I} \in \Cc\left(\Mm,\Lm\right)$. Similarly, the detection region $\Cc\left(\Mm,\Lm\right)$ is described, in this case, by the interval $\Cc\left({\mu},\epsilon\right)$ where $\mu\in [0,1)$ is the mean of the Bernoulli random variable describing the packet losses in the scalar case and $\epsilon\in [0,1]$ denotes the detection threshold set by the operator. 
Within this setting, the attack strategy is characterised by the attack design parameter $\alpha$. In view of this, substituting $\alpha$ as the control variable in (\ref{eq:basic}) reduces the optimisation problem to 
%
%The reader may notice similarities with the maximisation in this paper and the minimisation of the control law in~(\cite{MPC}). However, $\Mm^{\alpha}\neq \Mm$ and terms within $\Upsilon_{k|\mathcal{G}_k}^*$~(\cite{MPC}) do not cancel. At this stage it is assumed that all actuators for the system share a single communication channel, as in~\cite{1}. This initially simplifies the first attack construction. As a result of this $\bar{\Vs}^\alpha$ is a diagonal matrix with all elements equal. Therefore, $\Vs^\alpha_k=\alpha{\bf I} $ where $\alpha \in \R$, ${\bf I}\in \R^{Nn\times Nn}$, and $\alpha \in \Cc\left(\mu,\epsilon\right)$. Substituting this as the control variable in (\ref{eq:basic}) and $-\Gm_{\mathcal{G}_k}^{-1}\Fm X_k$ for $\Upsilon_{k|\mathcal{G}_k}$ yields
%%
\beqn
J^*_A\left(\Ac_k\right)&=&X_k^{\sf T}\left( \Qm +{\Delta}^\Phi \right)X_k + \tr\left(\Sigma_{\Xi}{\Delta}^{\Lambda}\right) \nonumber\\
&&\nqquad\nqquad +\!\!\!\! \max_{\alpha\in\Cc\left({\mu},\epsilon\right)}\!\!\left\{\!\Upsilon_{k|\mathcal{G}_k}^{\sf T} \alpha\left(\alpha{\Delta}^\Gamma\!\! +\! \left(1-\alpha\right)\left({\bf I} \odot{\Delta}^\Gamma \right)\!+\! {\Psi} - 2 \Gm_{\mathcal{G}_k}\right)\!\!\Upsilon_{k|\mathcal{G}_k}\right\}.\nonumber
\eeqn
%%
%\beqn
%J^*_A\left(\Ac_k\right)&=& X_k^{\sf T}\left( \Qm + {\Delta}^\Phi \right) X_k+\tr\left({\Delta}^{\Lambda} \Sigma_{\Xi}\right) \nonumber \\
%&&\nqquad\nqquad\nqquad + \max_{\bar{\Vs}^\alpha}\left\{ X_k^{\sf T}\Fm^{\sf T}\Gm_{\mathcal{G}_k}^{-1}\bar{\Vs}^\alpha \left( {\Delta}^\Gamma \bar{\Vs}^\alpha +{\Psi} +\left({\bf I}\odot {\Delta}^{\Gamma}\right)\left({\bf I}- \bar{\Vs}^{\alpha}\right)-2\Gm_{\mathcal{G}_k}\right)\Gm_{\mathcal{G}_k}^{-1}\Fm X_k\right\},\nonumber \\
%
%&&\nqquad\nqquad\nqquad =X_k^{\sf T}\left( \Qm +{\Delta}^\Phi \right)X_k + \tr\left(\Sigma_{\Xi}{\Delta}^{\Lambda}\right) \nonumber\\
%&&\nqquad\nqquad\nqquad + \max_{\alpha}\left\{ X_k^{\sf T} \Fm^{\sf T}\Gm_{\mathcal{G}_k}^{-1} \alpha\left(\alpha{\Delta}^\Gamma + \left(1-\alpha\right)\left({\bf I} \odot{\Delta}^\Gamma \right)+ {\Psi} - 2 \Gm_{\mathcal{G}_k}\right)\Gm_{\mathcal{G}_k}^{-1} \Fm X_k\right\}. \label{eq:f-UDP}
%\eeqn
%%
Note that the first two terms on the right hand side of the equation above are constants that do not depend on $\alpha$. Therefore, it is sufficient to maximise the last term. Substituting $-\Gm_{\mathcal{G}_k}^{-1}\Fm X_k$ for $\Upsilon_{k|\mathcal{G}_k}$ we write the term inside the maximisation as
\beqn\label{eq:f-UDP}
f\left(\alpha\right) &\triangleq& X_k^{\sf T} \Fm^{\sf T}\Gm_{\mathcal{G}_k}^{-1} \alpha\left(\alpha{\Delta}^\Gamma + \left(1-\alpha\right)\left({\bf I} \odot{\Delta}^\Gamma \right)+ {\Psi} \qquad\qquad  \right. \nonumber \\
&&\qquad \qquad \qquad \qquad \qquad\qquad  \left. - 2 \Gm_{\mathcal{G}_k}\right)\Gm_{\mathcal{G}_k}^{-1} \Fm X_k.
\eeqn
The function (\ref{eq:f-UDP}) is concave, convex, or linear in $\alpha$ depending on the system parameters, and therefore, the attacker has no control over the convexity of the cost function used for the attack construction. 
%
%For example, it is seen, via simulations, that the convexity of the function depends on the matrix ${{\bf A}}$ and $\Mm$.  %Move to numerical results?
%
However, the information set available to the attacker determines the convexity of the cost function, and therefore, the attacker is able to construct the optimal attack by solving (\ref{eq:f-UDP}) for any system parameters.
% 
%It could be argued that this is therefore irrelevant to the attacker, provided it is known whether the given realisation is convex and the attacker knows how to maximise the given function.
%
%The system parameters give rise to three different cases for function (\ref{eq:f-UDP}), specifically
%
In the following lemma we show that for the convex and linear systems the optimal attack construction is equivalent.
\begin{lem}\label{lem2}
	Let (\ref{eq:f-UDP}) be convex or linear in $\alpha$ over $\Cc\left(\mu,\epsilon\right)$. Then it's maximum is given by
	\vspace{-3mm}
	\beqn
	\max\{f(\alpha)\} =\max \left\{ f(\min\{\Cc\left(\mu,\epsilon\right)\}),f(\max\{\Cc\left(\mu,\epsilon\right)\})\vphantom{\Big |}\right\}.\nonumber
	\eeqn
\end{lem}
\begin{pf}
	Assume there is a maximum of (\ref{eq:f-UDP}), $f(a)$, such that $a \in \textnormal{Int}\{\Cc\left(\mu,\epsilon\right)\}$, we prove by contradiction that this is false, and therefore, the maximum is on the boundary of $\Cc\left(\mu,\epsilon\right)$. By the definition of convexity, for $\delta>0$ it holds that
	\vspace{-3mm}
	\beqn
	f(a) &\ge& \max\{f(a +\delta),f(a -\delta)\},\\
	f(a) &\ge& t f(a +\delta) + (1-t)f(a -\delta).
	\eeqn
It follows that $f(a)$ is greater than any point of the line connecting $f(a +\delta)$ and $f(a -\delta))$ however, this breaks the convexity assumption of (\ref{eq:f-UDP}), and therefore, the maximum is on the boundary. This concludes the proof. 
\hfill \ep
\end{pf}
%Add final description of a+\delta being in C?

%
When the function is concave there is a third maximising possibility, the case for which the global maximum of the function exists within the interval $\Cc\left(\mu,\epsilon\right)$. The following lemma describes this case. 
\begin{lem}\label{lem5}
	Let (\ref{eq:f-UDP}) be concave in $\alpha$ over $\Cc\left(\mu,\epsilon\right)$. Then the maximum of the function is given by
	\vspace{-3mm}
	\beqn
	\label{eq:concave_sol}
	\max\{f(\alpha)\} &=& \max\left\{ f(\min\{\Cc\left(\mu,\epsilon\right)\}),\vphantom{\Big |}\qquad\qquad\qquad\right. \nonumber \\
	 && \nqquad\nqquad  \left. f(\max\{\Cc\left(\mu,\epsilon\right)\}), f\left({\mathds 1}_{\Cc\left(\mu,\epsilon\right)}\left(\alpha_{\max}\right) \alpha_{\max}\right)\vphantom{\Big |}\right\}, 
	\eeqn
	where $ {\mathds 1}_\Bc\left(\alpha_{\max}\right)$ is the indicator function as a function of $\alpha_{\max}$ over the set $\Bc$ and $\alpha_{\max} \in\R$ is the global maximum of $f\left(\alpha\right)$.
\end{lem}
\begin{pf}
	In the concave case a global maximum exists, but is not necessarily within the interval $\Cc\left(\mu,\epsilon\right)$, and therefore, we restrict the domain to the safe operation region with the indicator function ${\mathds 1}_{\Cc\left(\mu,\epsilon\right)}\left(\alpha\right)$. The concavity of the function implies
	\vspace{-3mm}
	\beqn
	%f\left(\alpha\right) \!\!&\triangleq&\!\!\Upsilon^{* \sf T}_{k|\Gc_k} \alpha\left(\alpha{\Delta}^\Gamma + \left(1-\alpha\right)\left({\bf I} \odot{\Delta}^\Gamma \right)+ {\Psi} - 2 \Gm_{\mathcal{G}_k}\right) \Upsilon^{*}_{k|\Gc_k} ,\nonumber \\
	f^{\prime}(\alpha) &=& \Upsilon^{* \sf T}_{k|\Gc_k}\!\!\!\left(2\alpha{\Delta}^\Gamma\!\! +\! \left(1-2\alpha\right)\left({\bf I} \odot{\Delta}^\Gamma \right)\! + {\Psi}\! -\! 2 \Gm_{\mathcal{G}_k}\right)\!\! \Upsilon^{*}_{k|\Gc_k}\!,\ \quad \label{eq:firstderiv}\\
	f^{\prime\prime}(\alpha)&=& 2 \Upsilon^{* \sf T}_{k|\Gc_k}\left({\Delta}^\Gamma -\left({\bf I} \odot{\Delta}^\Gamma \right)\right) \Upsilon^{*}_{k|\Gc_k} < 0, \label{eq:secondderiv}
	\eeqn
	%
%	%
	where (\ref{eq:secondderiv}) follows from the strict concavity of (\ref{eq:f-UDP}). Setting (\ref{eq:firstderiv}) equal to zero gives
	\beqn
	&& \Upsilon^{* \sf T}_{k|\Gc_k}\left(2\alpha{\Delta}^\Gamma + \left(1-2\alpha\right)\left({\bf I} \odot{\Delta}^\Gamma \right) + {\Psi} - 2 \Gm_{\mathcal{G}_k}\right) \Upsilon^{*}_{k|\Gc_k} =0,\nonumber
\eeqn
which results in
\beqn
\label{eq:alpha_solve}
	&&\nqquad\nquad	2\alpha X_k^{\sf T} \Fm^{\sf T}\Gm_{\mathcal{G}_k}^{-1}\left({\Delta}^\Gamma -\left({\bf I} \odot{\Delta}^\Gamma \right)\right)\Gm_{\mathcal{G}_k}^{-1} \Fm X_k = \nonumber \\
	&& \qquad X_k^{\sf T} \Fm^{\sf T}\Gm_{\mathcal{G}_k}^{-1}\left(2\Gm_{\mathcal{G}_k}-{\Psi} -\left({\bf I} \odot{\Delta}^\Gamma \right)\right)\Gm_{\mathcal{G}_k}^{-1} \Fm X_k.
	\eeqn
It follows from the strict concavity of (\ref{eq:f-UDP}), as in (\ref{eq:secondderiv}), that 
\beqn
X_k^{\sf T} \Fm^{\sf T}\Gm_{\mathcal{G}_k}^{-1}\left({\Delta}^\Gamma -\left({\bf I} \odot{\Delta}^\Gamma \right)\right)\Gm_{\mathcal{G}_k}^{-1} \Fm X_k\ne0.
\eeqn
In view of this (\ref{eq:alpha_solve}) can be solved for $\alpha$ yielding
	\beqn
	\alpha_{\max} &=& %{{1}\over{2}} \underbrace{\left( X_k^{\sf T} \Fm^{\sf T}\Gm_{\mathcal{G}_k}^{-1}\left({\Delta}^\Gamma -\left({\bf I} \odot{\Delta}^\Gamma \right)\right)\Gm_{\mathcal{G}_k}^{-1} \Fm X_k \right)^{-1} }_{h^{-1}} \nonumber \\
	%&& \qquad \times \left(X_k^{\sf T} \Fm^{\sf T}\Gm_{\mathcal{G}_k}^{-1}\left(2\Gm_{\mathcal{G}_k}-{\Psi} -\left({\bf I} \odot{\Delta}^\Gamma \right)\right)\Gm_{\mathcal{G}_k}^{-1} \Fm X_k\right)\nonumber \\
	\!\!{{1}\over{2}}h^{-1}_\textnormal{UDP}\!\!\left(\!X_k^{\sf T} \Fm^{\sf T}\Gm_{\mathcal{G}_k}^{-1}\!\!\left(2\Gm_{\mathcal{G}_k}-{\Psi} -\left({\bf I} \odot{\Delta}^\Gamma \right)\right)\!\!\Gm_{\mathcal{G}_k}^{-1} \Fm X_k\!\right),\qquad \label{eq24}
	\eeqn
where $h_\textnormal{UDP}\eqdef\Upsilon^{* \sf T}_{k|\Gc_k}\left({\Delta}^\Gamma -\left({\bf I} \odot{\Delta}^\Gamma \right)\right) \Upsilon^{*}_{k|\Gc_k}$.
%	
%The $\alpha_{\max} $ above is the globally maximising $\alpha$ for the function $f(\cdot)$. 
%%
The global maximum is the solution when $\alpha_{\max} \in \Cc\left(\mu,\epsilon\right)$, i.e. the term $\alpha_{\max}\mathds{1}_{\Cc\left(\mu,\epsilon\right)}\left(\alpha_{\max}\right)$ in (\ref{eq:concave_sol}). When $\alpha_{\max}\notin\Cc\left(\mu,\epsilon\right)$ the solution follows as in the convex scenario by noticing that the inequality is strict and in the opposite direction. 
%
%
%The possibility of a globally maximising $\alpha$ becomes an interesting phenomenon in the case of detection constraints. This presents the possibility of a globally maximising $\alpha$ that is undetectable, this would be the best possible scenario for the attacker. 
%
Therefore, if $\alpha_{\max}\notin\Cc\left(\mu,\epsilon\right)$ the attack construction reverts to selecting the value of~$\alpha$ on the maximising boundary. 
For a concave function this is equivalent to finding the boundary that is closest to $\alpha_{\max}$. 
Let~$a,b\in\Cc\left(\mu,\epsilon\right)$ and assume $f(a)>f(b)$ and that $|a-\alpha_{\max}|<|b-\alpha_{\max}| $, then 
	\beqn
	f(b) &<& t f(a) + (1-t)f(\alpha_{\max}).
	\eeqn
However, this line segment lies above the function which contradicts the fact that this function is concave, and therefore, the maximising $\alpha$ is on the boundary that is closest to $\alpha_{\max}$. This concludes the proof. \ep
\end{pf}
Note that the $\alpha_{max}$ attack construction provides a globally optimal performance for the attacker from within the safe operation region. In fact, it also provides a lower probability of attack detection as it allows the attacker to operate away from the boundary. 

The following lemma highlights that an attack that minimises the cost of the operator is not achieved by setting $\alpha=1$. 
In the following we show that the optimal attack construction does not necessarily imply increasing the number of packet losses incurred by the operator. Indeed, there exist system parameters for which the optimal attack entails increasing the number of actuations. Whilst this might not be implementable in all attack scenarios, it is feasible to envision settings in which the attacker has full control of the actuation channel and can set the packet loss statistics at will. 
The following lemma captures this notion, namely, that the performance of the operator does not necessarily improve with the average number of received packets. Reiterating that the operator assumes a mean packet loss, and in doing so, creates an opportunity for the attacker to exploit the channel.
%
%% CHECK
\begin{lem}\label{lem1}
%	
%Let the convexity of (\ref{eq:f-UDP}) be unknown. The minimising choice of $\alpha$ is not equal to $\mu{\bf I}=\bar{\Vs}$ or ${1}$.
%
For any choice of system parameters it holds that
\be
\min_{a\in [0,1)} f(a)\leq \min\left\{f(1),f(\mu)\right\},
\ee
where $f$ is defined in (\ref{eq:f-UDP}).
\end{lem}
\begin{pf}
	Setting (\ref{eq:firstderiv}) equal to zero, and substituting in $\alpha = 1$ yields
	\beqn
	f^\prime(1) =\Upsilon_{k|\mathcal{G}_k}^{*\sf T} \left(2\left(\mathbf{I}-\bar{\Vs}\right){\Delta}^{\Gamma} - {\Psi} -\left(3\mathbf{I} -2\bar{\Vs}\right)\left({\bf I} \odot {\Delta}^{\Gamma}\right)  \right)\Upsilon_{k|\mathcal{G}_k}^*. \nonumber 
	\eeqn	
	For this to be a minimising solution it needs to hold that, ${\Psi}$ is equal to $2\left({\bf I}-\bar{\Vs}\right){\Delta}^{\Gamma}-\left(3{\bf I}-2\bar{\Vs}\right)\left({\bf I} \odot {\Delta}^{\Gamma}\right)$. Due to ${\Psi} $ being a diagonal matrix and the structure of ${\Delta}^\Gamma$ it is only possible for this equality to hold in a system with ${\Am}={\bf 0}$ and a diagonal $\Bm$. In this scenario, ${\Delta}^{\Gamma}=\left({\bf I} \odot {\Delta}^{\Gamma}\right)$, this results in, ${\Psi}=-\left({\bf I} \odot {\Delta}^\Gamma\right)$. By assumption ${\Psi}\succ 0$, however, it is shown in Lemma \ref{lem4} that $\left({\bf I} \odot {\Delta}^\Gamma\right)\succ 0$ which is a contradiction, and therefore, $f^\prime(\alpha=1)\ne 0$ and thus $\alpha=1$ is not a minimising solution. Substituting $\alpha{\bf I}=\mu{\bf I} = {\bar{\Vs}}$ in  (\ref{eq:firstderiv}) results in	%
	\beqn
	f^\prime(\alpha) &=&\Upsilon_{k|\mathcal{G}_k}^{*\sf T} \left(2\bar{\Vs}{\Delta}^\Gamma + \left(\mathbf{I}-2\bar{\Vs}\right)\left({\bf I} \odot{\Delta}^\Gamma \right)+ {\Psi} - 2 \Gm_{\mathcal{G}_k}\right)\Upsilon_{k|\mathcal{G}_k}^{*},\nonumber\\
	&=&-X_k^{\sf T} \Fm^{\sf T}\Gm_{\mathcal{G}_k}^{-1} \left( {\Psi} +\left({\bf I} \odot {\Delta}^{\Gamma}\right)\right)\Gm_{\mathcal{G}_k}^{-1} \Fm X_k .\label{eq49}
	\eeqn
	This is not equal to $0$ due to ${\Psi}+\left({\bf I} \odot {\Delta}^\Gamma\right)\succ 0$. Therefore,~(\ref{eq49}) is strictly negative and not a minimising solution. This concludes the proof. \ep
\end{pf}
%
%%CHECK

%% Turn into THEOREM - Same TCP
\begin{thm}\label{thm1}
Let $\Ac_k=\left\{{{\bf A}},{{\bf B}},\Sigma_{W},\bar{\nu}, \Omega, {\Psi}, \Gc_k \right\}$ be the information set available to construct the attack, then the optimal mean packet loss probability for an IID attack is given by 
	\beqn
	\alpha^*_{\textnormal{UDP}} &=& \max\left\{f\left(\min\{\Cc\left(\mu,\epsilon\right)\}\right),\vphantom{\Big |}f\left(\max\{\Cc\left(\mu,\epsilon\right)\}\right),\nonumber \right.\\
	&&\left.\qquad\qquad\qquad\qquad\qquad  f\left({\mathds 1}_{\Cc\left(\mu,\epsilon\right)}\left(\alpha_{\max}\right)\alpha_{\max}\right)\vphantom{\Big |}\right\},\nonumber 
	\eeqn
where 
	\beqn
	f\left(a\right) &\triangleq& X_k^{\sf T} \Fm^{\sf T}\Gm_{\mathcal{G}_k}^{-1} a\left(a{\Delta}^\Gamma + \left(1-a\right)\left({\bf I} \odot{\Delta}^\Gamma \right)+ {\Psi} \qquad\qquad  \right. \nonumber \\
	&&\qquad \qquad \qquad \qquad \qquad\qquad  \left. - 2 \Gm_{\mathcal{G}_k}\right)\Gm_{\mathcal{G}_k}^{-1} \Fm X_k.\nonumber 
	\eeqn
%	
%	
%	
%	for a system operating under a UDP-like communication protocol is seen in Lemma~\ref{lem2} and Lemma~\ref{lem5}. It is seen in Lemma~\ref{lem2}  how to choose the optimal $\alpha$ for a convex/linear $f(\alpha)$ and Lemma~\ref{lem5} shows the optimal $\alpha$ for a convex $f(\alpha)$. Therefore, by combining the two lemmas it is seen that the optimal choice of $\alpha$ that maximises $f(\alpha)$ is
%	%
%	\beqn
%	\alpha^*_{\textnormal{UDP}} &=& \max\left\{f\left(\min\{\Cc\left(\mu,\epsilon\right)\}\right),\vphantom{\Big |}f\left(\max\{\Cc\left(\mu,\epsilon\right)\}\right),\nonumber \right.\\
%	&&\left.\qquad\qquad\qquad\qquad\qquad\qquad\qquad  f\left({\mathds 1}_{\Cc\left(\mu,\epsilon\right)}\alpha_{\max}\right)\vphantom{\Big |}\right\}.\nonumber 
%	\eeqn
%	Which is equivalent to checking the detection boundaries and the function maximum.

\end{thm}

\begin{pf}
The result follows from the application of Lemma 1 for the convex and linear cases, Lemma 2 for the concave case, and by noticing that the set of solutions for the convex and linear cases is a subset of of the set of solutions of the concave case. This concludes the proof. \ep
\end{pf}	
\subsection{TCP-like Protocol}
The optimal attack strategy  for the TCP-like protocol is the solution to the optimisation problem
\begin{subequations}\label{eq:TCP-opt}
	\beqn
	&&\max_{\bar{\Vs}^\alpha}\; \;J_A\left(\Ac_k\right), \\
	&& \mbox{s.t.}\;\quad \Mm^\alpha \in \Cc\left(\Mm,\Lm\right).
	\eeqn
\end{subequations}
Note that in this case the information set $\Ac_k$ contains the realisations of the packet losses as given in $\Fc_k$. For that reason, the optimisation problem differs from that in (\ref{eq:UDP-opt}) in that the cost function exhibits a different structure induced by the conditioning of the previous packet loss realisations. 
From (\ref{eq11}) and Lemma 1 in \cite{MPC} with substitution of the optimal control law under the TCP-like protocol in (\ref{eq:Upsilon}) yields
%
%\beqn
%\nquad J^*(\Ac_k)&=& X_k^{\sf T}\left( \Qm + {\Delta}^\Phi \right) X_k+\tr\left({\Delta}^{\Lambda} \Sigma_{\Xi}\right) \nonumber \\
%&& \!\!\!\!\!\!\!\!\!+ \max_{\bar{\Vs}^\alpha}\left\{ \Upsilon_{k|\mathcal{F}_k}^{*{\sf T}}\bar{\Vs}^\alpha \left( 2\Fm X_k + \left({\Delta}^\Gamma \bar{\Vs}^\alpha +{\Psi} \right)\Upsilon_{k|\mathcal{F}_k}^*\right)\right\}\!\!,\label{eq28} \\
%\eeqn
%
%
\beqn
&& J^*(\Ac_k)= 
%	X_k^{\sf T}\left( \Qm + {\Delta}^\Phi \right) X_k+\tr\left({\Delta}^{\Lambda} \Sigma_{\Xi}\right) \nonumber \\
%	&& + \max_{\bar{\Vs}^\alpha}\left\{ X_k^{\sf T}\Fm^{\sf T} \Gm_{\mathcal{F}_k}^{-1}\bar{\Vs}^\alpha \left( -2\Fm X_k + \left({\Delta}^\Gamma \bar{\Vs}^\alpha +{\Psi} \right)\Gm_{\mathcal{F}_k}^{-1}\Fm X_k\right)\right\},\nonumber \\
X_k^{\sf T}\left( \Qm + {\Delta}^\Phi \right) X_k+\tr\left({\Delta}^{\Lambda} \Sigma_{\Xi}\right) \nonumber \\
&& +\max_{\bar{\Vs}^\alpha}\left\{ X_k^{\sf T}\Fm^{\sf T} \Gm_{\mathcal{F}_k}^{-1}\bar{\Vs}^\alpha \left( {\Delta}^\Gamma \bar{\Vs}^\alpha   +{\Psi} -2\Gm_{\mathcal{F}_k} \right)\Gm_{\mathcal{F}_k}^{-1}\Fm X_k\right\}\!, \quad\label{eq:31}
%&& \mbox{s.t.}\;\quad \Mm^\alpha \in \Cc\left(\Mm,\Lm\right). \nonumber 
\eeqn
where the maximisation is subject to $\Mm^\alpha \in \Cc\left(\Mm,\Lm\right)$. As with the UDP-like protocol attack construction, it is assumed without loss of generality, that all actuators share a single communication channel~(\cite{1}). Therefore, $\bar{\Vs}^\alpha=\alpha{\bf I}$. Noting that the first two terms in (\ref{eq:31}) do not depend on $\bar{\Vs}^\alpha$ and that $\Gm_{\mathcal{F}_k}=\left({\Delta}^\Gamma\bar{\Vs}+{\Psi}\right)$  as shown in (\cite{MPC}), then the term inside the maximisation can be rewritten as
%, and therefore, (\ref{eq:31}) is reformulated as the functional
%
%\beqn
%g\left(\alpha\right)&\triangleq& -X_k^{\sf T} \Fm^{\sf T}\Gm_{\mathcal{F}_k}^{-1}\bar{\Vs}^\alpha \left( {\Delta}^{\Gamma}\left(2\bar{\Vs}-\bar{\Vs}^\alpha \right) + {\Psi} \right)\Gm_{\mathcal{F}_k}^{-1}\Fm X_k.\ \ \label{eq:g-TCP-alpha}
%\eeqn
%
% and as a result (\ref{eq:g-TCP-alpha}) is
%
\beqn
g\left(\alpha\right)&\triangleq& -X_k^{\sf T} \Fm^{\sf T}\Gm_{\mathcal{F}_k}^{-1}\alpha \left( {\Delta}^{\Gamma}\left(2\bar{\Vs}-\alpha\mathbf{I} \right) + {\Psi} \right)\Gm_{\mathcal{F}_k}^{-1}\Fm X_k. \label{eq:g-TCP}
\eeqn
Differentiating (\ref{eq:g-TCP}) results in
\beqn
g^{\prime}\left(\alpha\right) &= & -X_k^{\sf T} \Fm^{\sf T}\Gm_{\mathcal{F}_k}^{-1}\left( {\Delta}^{\Gamma}\left(2\bar{\Vs}-2\alpha\mathbf{I} \right) + {\Psi} \right)\Gm_{\mathcal{F}_k}^{-1}\Fm X_k,\label{eq32} \\
g^{\prime\prime}\left(\alpha\right) &= & 2 X_k^{\sf T} \Fm^{\sf T}\Gm_{\mathcal{F}_k}^{-1}{\Delta}^{\Gamma}\Gm_{\mathcal{F}_k}^{-1}\Fm X_k.\label{eq:second-deriv-TCP}
\eeqn

\begin{lem}\label{lem4}
	Let the pair $({{\bf A}},{\bf B})$ be reachable and the state penalty matrix $\Omega$ be positive definite. Then the function defined by (\ref{eq:g-TCP}) is convex in $\alpha$ over $\Cc\left(\mu,\epsilon\right)$ almost surely.
	\end{lem}
	\begin{pf}
	It follows from (\ref{eq:second-deriv-TCP}) that if ${\Delta}^{\Gamma}\succ 0$ and $X_k\neq{\bf 0}$ then (\ref{eq:second-deriv-TCP}) is strictly greater than zero. Therefore, (\ref{eq:g-TCP}) is convex in $\alpha$ over $\Cc\left(\mu,\epsilon\right)$. It is shown in~\cite[p.225, 10.31(c)]{6} that when $\textnormal{rank}\left(\Gamma\right)=\max\{Nn,Nm\}$ and $\Omega$ is positive definite then ${\Delta}^\Gamma\succ 0$. Since $ ({{\bf A}},{\bf B})$ is a reachable pair then $\textnormal{rank}\left[ \Bm, \Am\Bm,\ldots,\Am^{N-1}\Bm\right]=\max\{n,m\}$. Therefore, due to the triangular structure of $\Gamma$ we have that $\textnormal{rank}\left(\Gamma\right)=\max\{Nn,Nm\}$. Under these assumptions (\ref{eq:second-deriv-TCP}) is convex in $\alpha$ over $\Cc\left(\mu,\epsilon\right)$ when $X_k\neq0$, which holds with probability $1$.
	%
	%In order for ${\Delta}^{\Gamma}$ to be positive definite For this to hold the state penalty matrix must be positive definite, $\Omega \succ 0 $, and the pair $({{\bf A}},{\bf B})$ must be reachable. 
	This concludes the proof.\ep
	\end{pf}
\begin{thm}\label{thm2} % Copy format of Theorem 4 (UDP case)
	Consider a system operating with a TCP-like protocol.
	\beqn
	g\left(\alpha\right)&\triangleq& -X_k^{\sf T} \Fm^{\sf T}\Gm_{\mathcal{F}_k}^{-1}\alpha \left( {\Delta}^{\Gamma}\left(2\bar{\Vs}-\alpha{\bf I} \right) + {\Psi} \right)\Gm_{\mathcal{F}_k}^{-1}\Fm X_k,\nonumber 
	\eeqn
	 Therefore, the optimal choice of $\alpha$ is
	\beqn
	\alpha^*_\textnormal{TCP} = \max\left\{g\left(\min\{\Cc\left(\mu,\epsilon\right)\}\right),g\left(\max\{\Cc\left(\mu,\epsilon\right)\}\right)\vphantom{\Big |}\right\}.\nonumber 
	\eeqn
\end{thm}
\begin{pf}
	Note from Lemma \ref{lem4} that $g(\alpha)$ is convex therefore, as shown in Lemma \ref{lem1}, $\alpha^{*}_{TCP}$ is known to be on the boundary. This concludes the proof.
	\end{pf}
Note that due to the convexity of (\ref{eq:g-TCP}) the solution of (\ref{eq32}) results in the minimising value of $\alpha$, which interestingly is not $\alpha{\bf I}=\bar{\Vs}$, or $\alpha = 1$ but instead is given by
\beqn
\alpha_{\min}={{1}\over{2}}h_\textnormal{TCP}^{-1} \left(X_k^{\sf T} \Fm^{\sf T}\Gm_{\mathcal{G}_k}^{-1}\!\!\left(2{\Delta}^\Gamma\bar{\Vs} + {\Psi}\right)\Gm_{\mathcal{G}_k}^{-1} \Fm X_k\right),
\eeqn
where $h_{TCP}= X_k^{\sf T} \Fm^{\sf T}\Gm_{\mathcal{F}_k}^{-1}{\Delta}^{\Gamma}\Gm_{\mathcal{F}_k}^{-1}\Fm X_k > 0$. When considering the TCP-like protocol without detection constraints, additional insight can be obtained by analysing the attack construction
\beqn
g\left(1\right)&=&X_k^{\sf T} \Fm^{\sf T}\Gm_{\mathcal{F}_k}^{-1} \left( {\Delta}^{\Gamma}\left({\bf I}- 2\bar{\Vs}\right) - {\Psi} \right)\Gm_{\mathcal{F}_k}^{-1}\Fm X_k, \ g\left(0\right)= 0.\ \ \ \ \label{eq:g1}
\eeqn
From (\ref{eq:g1}), if ${\Delta}^{\Gamma}\left({\bf I}- 2\bar{\Vs}\right) \succ {\Psi} $ the maximising value of $\alpha$ is $1$ or $0$. 
%
%This term can be analysed to give an idea of what this means for the system. 
%
The ${\Delta}^{\Gamma}\left({\bf I}- 2\bar{\Vs}\right) $ term is the state penalty matrix $\Omega$ weighted by the reachability of the system and the packet loss probability. 
The terms in (\ref{eq:g1}) capture the average impact of actuation in the cost reduction with respect to the input penalty matrix ${\Psi}$. 
%This term relates to the average ability of an actuation to reduce the cost, and this is being compared to the input penalty matrix, ${\Psi}$. 
%%
Therefore, the optimal attack is $1$ when the average cost increase per actuation is greater than the average penalty induced by the actuation. 
As a result, for a system with a high probability of packet loss that penalises state error more than actuation, the optimal attack strategy is to allow perfect communication, i.e. all packets are received by the plant.
Additionally, for $\bar{\Vs}\succ{{1}\over{2}}{\bf I} $ the optimal attack strategy is $\bar{\Vs}={\bf 0}$. That being the case, for a system with a low probability of packet losses the operator could simplify their detection criteria to a one-sided test.
\section{Cost Increase Analysis}
In this section we evaluate the cost increase induced by the optimal IID attack by comparing the expected cost when an attack is present to the expected cost when no attack is present, i.e.  $\mathds{E}\left[J^*_A(\Ac_k)\right] - \mathds{E}\left[ J^*(\Ic_k)\right]$. The expected cost increase of the three attack strategies are studied separately. The analysis is carried out for the case $\Cc\left(\mu,\epsilon\right)=[0,1)$, i.e. the extreme cases of the average attack packet drop.  
Note that there is no loss of generality as the case with detection constraints can be analysed following the same approach with the appropriate scaling. Additionally, when considering the different cases of $\alpha^*$ it should be noted that for a given set of detection parameters $\alpha^*$ is unique. Since, the detection parameters, $\mu,\epsilon$, are not fixed. The region $[0,1)$ is continuous with respect to $\alpha^*$.
%%
%In this section we analyse the cost increase induced by the optimal attack constructions presented. In order to assess this the attacker calculates the expected induced cost from a construction and compares it with the operators expected cost. These attack constructions are probabilistic attacks and on average increase the operators cost, this does not mean that every realisation of the attack is above the expected cost of the operator.
%
\newsavebox{\smlA}% Box to store smallmatrix content
\savebox{\smlA}{$\left(\begin{smallmatrix} 1.03&0.005\\0.35&0.5\end{smallmatrix}\right)$}
\newsavebox{\smlB}% Box to store smallmatrix content
\savebox{\smlB}{$\left(\begin{smallmatrix}1&0\\0&1\end{smallmatrix}\right)$}
\newsavebox{\smlV}% Box to store smallmatrix content
\savebox{\smlV}{$\left(\begin{smallmatrix}0.7&0\\0&0.01\end{smallmatrix}\right)$}
%
%
%\begin{figure}[!t]						%%TCP Scalar!
%	\captionsetup{justification=centering,margin=2cm,width=\linewidth}
%	\includegraphics[width=\linewidth]{Figures/Av_Attack_S.eps}
%	%		\includegraphics[width=\linewidth]{Figures/Av_Attack.eps}
%	\caption{System with TCP-like protocol with $ {\bf A} =\usebox{\smlA}$, $\Mm=0.7$, and $\epsilon=0.1$.}\label{fig3}
%	%	\includegraphics[draft,width=\textwidth]{Doubl.eps}
%\end{figure}
%
\subsection{UDP-like Cost Analysis}
%
%The aim of the attacker is to increase the expected cost of the system, i.e.  $\mathds{E}\left[J^*_A(\Ac_k)\right] \ge \mathds{E}\left[ J^*(\Ic_k)\right]>0$. The expected cost increase of the three attack strategies must all be characterised separately. The following section considers the whole space for characterisation, $\Cc\left(\mu,\epsilon\right)=[0,1)$, any attack with detection constraints will be a scaled version of the following. For a given system only one strategy is optimal, therefore the cost increase of that strategy is greater than all others for that given system.
%%
%The attacker is increasing the expected cost of the system, therefore, $\mathds{E}\left[J^*_A(\Ac_k)\right] \ge \mathds{E}\left[ J^*(\Gc_k)\right]>0$. The expected cost increase of the three attack strategies must all be characterised separately. The following section considers the whole space for characterisation, $\Cc\left(\mu,\epsilon\right)=[0,1)$, any attack with detection constraints will be a scaled version of the following. For a given system only one strategy is optimal, therefore the cost increase of that strategy is greater than all others for that given system.
%
\subsubsection{Attack performance when $\alpha^*\rightarrow0$.}
 %$\alpha*=\alpha^*\rightarrow0\rightarrow0$ \\
We first analyse the case when the attacker losses all the packets and induces the cost 
\be
\mathds{E}\left[J^0_A(\Ac_k)\right]\eqdef \lim_{\alpha^*\rightarrow 0}\mathds{E}\left[J_A(\Ac_k)\right].
\ee
The expected cost 
%for the operator when there is not attack and 
when there is an attack is given by
%
%\be
%\mathds{E}\left[J^*(\Gc_k)\right]=\tr\left(\Sigma_{X}\left(\Qm + {\Delta}^\Phi\right) + \Sigma_{\Xi}{\Delta}^{\Lambda}\right) - X_k^{\sf T} \Fm^{\sf T} \Gm_{\mathcal{G}_k}^{-1}\bar{\Vs} \Fm X_k,
%\ee
%and
\be
\mathds{E}\left[J^0_A(\Ac_k)\right]= \tr\left(\Sigma_{X}\left(\Qm + {\Delta}^\Phi\right) + \Sigma_{\Xi}{\Delta}^{\Lambda}\right)+ \max\{f(\alpha)\}.
\ee
Since (\ref{eq:f-UDP}) is continuous in $\alpha$ we have that $\alpha^*\rightarrow0$ implies $f\left(\alpha^*\right)\rightarrow0$, and therefore, the cost increase is
\be
\mathds{E}\left[J^0_A(\Ac_k)\right]-\mathds{E}\left[J^*({\mathcal{G}_k})\right]\
= X_k^{\sf T} \Fm^{\sf T} \Gm_{\mathcal{G}_k}^{-1}\bar{\Vs} \Fm X_k >0.\label{eq39}
\ee
%\alert{
Note that the $\alpha^*\rightarrow0$ attack strategy forces the system into open loop, and therefore, the expected cost increase coincides with the expected cost reduction introduced by the controller when there is no attack present in the communication channel.
%}
%\alert{(IE: This sentence might be a bit iffy. Please check if the control lingo is ok.)}
%\begin{figure}[!t] 						%% UDP Scalar!
%	\captionsetup{justification=centering,margin=2cm,width=\linewidth}
%	\includegraphics[width=\linewidth]{Figures/Av_Attack_UDP_S.eps}
%	\caption{System with UDP-like protocol with ${\bf A} =\usebox{\smlA}$, $\Mm=0.7$, and $\epsilon=0.1$.}\label{fig3a}
%	%	\includegraphics[draft,width=\textwidth]{Doubl.eps}
%\end{figure}
%
\subsubsection{Attack performance when $\alpha^*=1$.}
In this case, the attacker allows successful reception of all packets, i.e. the actuation communication channel is perfect. Surprisingly, there exist systems for which the cost increase, given by
\be
\mathds{E}\left[J^1_A(\Ac_k)\right]\eqdef \mathds{E}\left[J_A(\Ac_k)\right]\Big |_{\alpha^*=1},
\ee
is positive despite the fact that the communication channel of the operator improves. Evaluation of (\ref{eq:f-UDP}) with perfect communication results in
 
%The case when attacker allows more packets through than the operator expects is an optimal attack strategy for certain system parameters as shown earlier.

%In this setting the,  

%The $\alpha^*=1$ attack construction is not the most intuitive attack strategy as it allows more actuations through to the actuators than expected. This could be seen as `upgrading' the actuation channel by boosting the controllers transmitted signal.
%
\beqn
%J^*(\Gc_k)&=& X_k^{\sf T}\left( \Qm +{\Delta}^\Phi \right)X_k + \tr\left(\Sigma_{\Xi}{\Delta}^{\Lambda}\right) - X_k^{\sf T} \Fm^{\sf T} \Gm_{\mathcal{G}_k}^{-1}\bar{\Vs} \Fm X_k \\
\mathds{E}\left[J^1(\Ac_k)\right]&=& \tr\left(\Sigma_{X}\left(\Qm + {\Delta}^\Phi\right) + \Sigma_{\Xi}{\Delta}^{\Lambda}\right)\nonumber \\
&&+ X_k^{\sf T} \Fm^{\sf T}\Gm_{\mathcal{G}_k}^{-1} \left({\bf I}-2\bar{\Vs}\right)\left({\Delta}^{\Gamma} - \left({\bf I} \odot {\Delta}^{\Gamma}\right)\right)\Gm_{\mathcal{G}_k}^{-1} \Fm X_k \nonumber \\
&& -X_k^{\sf T} \Fm^{\sf T}\Gm_{\mathcal{G}_k}^{-1} \left(\left({\bf I} \odot {\Delta}^{\Gamma}\right)+ {\Psi}  \right)\Gm_{\mathcal{G}_k}^{-1} \Fm X_k.
\eeqn
Unlike the $\alpha^*\rightarrow0$ case, the $\alpha^*=1$ construction does not guarantee an increase in cost for every system. In fact, the cost only increases when
\beqn
&&X_k^{\sf T} \Fm^{\sf T}\Gm_{\mathcal{G}_k}^{-1} \left({\bf I}-2\bar{\Vs}\right)\left({\Delta}^{\Gamma} - \left({\bf I} \odot {\Delta}^{\Gamma}\right)\right)\Gm_{\mathcal{G}_k}^{-1} \Fm X_k\ \ge \nonumber \\ 
&&\qquad\qquad X_k^{\sf T} \Fm^{\sf T}\Gm_{\mathcal{G}_k}^{-1} \left(\left({\bf I} \odot {\Delta}^{\Gamma}\right)+ {\Psi}  \right)\Gm_{\mathcal{G}_k}^{-1} \Fm X_k> 0. \label{eq44}
\eeqn
However, all variables that determine (\ref{eq44}) are system parameters known by the attacker, and therefore, the attacker decides the optimal attack strategy accordingly. The expected cost increase is
\beqn
&&\mathds{E}\left[J^1_A(\Ac_k)\right]-\mathds{E}\left[J^*(\Gc_k)\right]\nonumber \\
&&= X_k^{\sf T} \Fm^{\sf T} \Gm_{\mathcal{G}_k}^{-1} \left({\Delta}^\Gamma + {\Psi} \right)\Gm_{\mathcal{G}_k}^{-1} \Fm X_k +\left(\bar{\Vs}-2{\bf I}\right)X_k^{\sf T} \Fm^{\sf T} \Gm_{\mathcal{G}_k}^{-1} \Fm X_k. \nonumber 
\eeqn 
\subsubsection{Attacker performance when $\alpha^* ={\mathds{1}_{\Cc\left(\mu,\epsilon\right)}\left(\alpha_{\max}\right)}\alpha_{\max}$.}
We tackle next the introduction of a general detection constraint.
%The prospect of being able to achieve a global maximum in the cost from an attack strategy is quite intriguing but in doing so requires further analysis. 
In this case, the expected cost for the attacker is
\beqn
%J^*(\Gc_k)&=& X_k^{\sf T}\left( \Qm +{\Delta}^\Phi \right)X_k + \tr\left(\Sigma_{\Xi}{\Delta}^{\Lambda}\right) - X_k^{\sf T} \Fm^{\sf T} \Gm_{\mathcal{G}_k}^{-1}\bar{\Vs} \Fm X_k \\
%
%
\mathds{E}\left[J^{\alpha_{\max}}(\Ac_k)\right]&\eqdef& \mathds{E}\left[J_A(\Ac_k)\right]\Big |_{\alpha^*=\alpha_{\textnormal{max}}}\nonumber\\
&=&\tr\left(\Sigma_{X}\left(\Qm + {\Delta}^\Phi\right) + \Sigma_{\Xi}{\Delta}^{\Lambda}\right) + f(\alpha_{\max}).\nonumber 
\eeqn
Algebraic maniupulation of $f(\alpha_{\max})$ and substituting (\ref{eq24}) yields:
%
%\beqn
%f\left(\alpha_{\max}\right)&=&X_k^{\sf T} \Fm^{\sf T} \Gm_{\mathcal{G}_k}^{-1}\left(\alpha_{\max}\left( {\Delta}^\Gamma \alpha_{\max} + \left(1-\alpha_{\max}\right)\left({\bf I} \odot {\Delta}^\Gamma\right) \right.\right. \nonumber \\
%&&  \qquad\qquad\left.\left.  +{\Psi}  - 2\Gm_{\mathcal{G}_k}\right)\right)\Gm_{\mathcal{G}_k}^{-1} \Fm X_k \nonumber \\
%%
%%
%%&=& X_k^{\sf T} \Fm^{\sf T} \Gm_{\mathcal{G}_k}^{-1}\alpha_{\max}\left( \alpha_{\max} \left({\Delta}^\Gamma -\left({\bf I} \odot {\Delta}^\Gamma\right)\right)\right)\Gm_{\mathcal{G}_k}^{-1} \Fm X_k \nonumber \\
%%&& \qquad + X_k^{\sf T} \Fm^{\sf T} \Gm_{\mathcal{G}_k}^{-1}\alpha_{\max}\left(\left({\bf I} \odot {\Delta}^\Gamma\right) +{\Psi} - 2\Gm_{\mathcal{G}_k}\right)\Gm_{\mathcal{G}_k}^{-1} \Fm X_k \nonumber \\
%%%
%%
%%&=&\alpha_{\max}^{2} h_\textnormal{UDP} - \alpha_{\max} X_k^{\sf T} \Fm^{\sf T} \Gm_{\mathcal{G}_k}^{-1}\left( 2\Gm_{\mathcal{G}_k} -\left({\bf I} \odot {\Delta}^\Gamma\right) -{\Psi}\right)\Gm_{\mathcal{G}_k}^{-1} \Fm X_k \nonumber \\
%%
%%
%&=&\!-\!{{1}\over{4}} \alpha_{\max}X_k^{\sf T} \Fm^{\sf T} \Gm_{\mathcal{G}_k}^{-1}\!\left( 2\Gm_{\mathcal{G}_k}\!\!-\left({\bf I} \odot {\Delta}^\Gamma\right) -{\Psi}\right)\!\Gm_{\mathcal{G}_k}^{-1} \Fm X_k. \nonumber 
%\eeqn
%%
%Substituting $\alpha_{\max}$ as given in (\ref{eq24}) yields:
%
\beqn
 f(\alpha_{\max})\!\!= \!\!\frac{ h_\textnormal{UDP}^{-1} }{4}%{{1}\over{4}} h_\textnormal{UDP}^{-1} %\nonumber \\
\left(X_k^{\sf T} \Fm^{\sf T}\Gm_{\mathcal{G}_k}^{-1}\left(2\Gm_{\mathcal{G}_k}-{\Psi} -\left({\bf I} \odot{\Delta}^\Gamma \right)\right)\Gm_{\mathcal{G}_k}^{-1} \Fm X_k\right)^2\!\!,\nonumber
\eeqn

where the inequality comes from the fact that $f$ is concave when $\alpha_{\textnormal{max}}$ is a feasible optimal attack strategy. The resulting cost increase is 
\beqn
%\!\!\!\!\Delta\mathds{E}\left[J^{\alpha_{\max}}\left(\Gc_k\right)\right]&=&
\mathds{E}\left[J^{\alpha_{\max}}\left(\Ac_k\right)\right] - J^*(\Gc_k)
&=&X_k^{\sf T} \Fm^{\sf T} \Gm_{\mathcal{G}_k}^{-1}\bar{\Vs} \Fm X_k \nonumber \\
&& \nqquad\nqquad\nqquad\nqquad \nqquad\nqquad \ \ \ + {{1}\over{4}} h_\textnormal{UDP}^{-1}\left(X_k^{\sf T} \Fm^{\sf T}\Gm_{\mathcal{G}_k}^{-1}\left(2\Gm_{\mathcal{G}_k}-{\Psi} -\left({\bf I} \odot{\Delta}^\Gamma \right)\right)\Gm_{\mathcal{G}_k}^{-1} \Fm X_k\right)^2 > 0. \nonumber
\eeqn
Note that the inequality is strict, i.e. the attack guarantees a performance loss of the operator. As mentioned previously this attack strategy is only feasible when $\alpha_{\max}\in\Cc\left(\mu,\epsilon\right)$ and (\ref{eq:f-UDP}) is concave. Additionally, $\mathds{E}\left[J^0\left(\Ac_k\right)\right]-\mathds{E}\left[J^*\left(\Gc_k\right)\right]$ is a upper bounded by the cost increase induced by the $\alpha^* ={\mathds{1}_{\Cc\left(\mu,\epsilon\right)}\left(\alpha_{\max}\right)}\alpha_{\max}$ strategy.
%
%\begin{figure}[!t] 						%% TCP !
%	\captionsetup{justification=centering,margin=2cm,width=\linewidth}
%	\includegraphics[width=\linewidth]{Figures/Av_Attack.eps}
%	\caption{System with TCP-like protocol with ${\bf A} =\usebox{\smlA}$, $\Mm=\usebox{\smlV}$, and $\Lm=0.1 {\bf I}$.}\label{fig4}
%\end{figure}

\subsection{TCP-like Cost Analysis}
The cost increase analysis for TCP-like protocols contains only two attack strategies. The analysis is again performed on the $\Cc\left(\mu,\epsilon\right)=[0,1)$ interval.
%
%\begin{figure}[!t]						%% UDP !
%	\captionsetup{justification=centering,margin=2cm,width=\linewidth}			
%	\includegraphics[width=\linewidth]{Figures/Av_Attack_UDP.eps}
%	\caption{System with UDP-like protocol with ${\bf A} =\usebox{\smlA}$, $\Mm=\usebox{\smlV}$, and $\Lm=0.1{\bf I}.$}\label{fig4a}
%\end{figure}
%
\subsubsection{Attacker performance when $\alpha^*\rightarrow0$.}
For the attack construction that forces the  system into open loop the expected cost is given by
\beqn
\hfill\mathds{E}\left[J^0\left(\Ac_k\right)\right]&=& \tr\left(\Sigma_{X}\left(\Qm + {\Delta}^\Phi\right) + \Sigma_{\Xi}{\Delta}^{\Lambda}\right)+ \lim_{\alpha^*\rightarrow 0}g(\alpha^*). \nonumber 
\eeqn
Since (\ref{eq:g1}) is continuous in $\alpha$ we have that $\alpha^*\rightarrow0$ implies $g(\alpha^*)\rightarrow0$. Therefore, the expected cost increase is:
\be
\mathds{E}\left[J^0\left(\Ac_k\right)\right]-\mathds{E}\left[J^*\left(\Fc_k\right)\right]%\nonumber \\
= X_k^{\sf T} \Fm^{\sf T} \Gm_{\mathcal{F}_k}^{-1}\bar{\Vs} \Fm X_k >0. \nonumber
\ee
As with the UDP-like protocol, by implementing the $\alpha^*\rightarrow0$ attack strategy the attacker forces the system into open loop. Note that the cost increase for the TCP-like protocol and the UDP-like protocol under the $\alpha^*\rightarrow0$ strategy differ only in the $\Gm_{\Ic_k}$ designed by the controller, i.e. on the available information.
\subsubsection{Attacker performance when $\alpha^*=1$.}
In the TCP case, the attack that provides a perfect communication channel induces an expected cost given by
\beqn
%\mathds{E}\left[J^*\left(\Fc_k\right)\right]&=&\tr\left(\Sigma_{X}\left(\Qm + {\Delta}^\Phi\right) + \Sigma_{\Xi}{\Delta}^{\Lambda}\right) - X_k^{\sf T} \Fm^{\sf T} \Gm_{\mathcal{F}_k}^{-1}\bar{\Vs} \Fm X_k \nonumber \\
\mathds{E}\left[J^1\left(\Ac_k\right)\right]&=& \tr\left(\Sigma_{X}\left(\Qm + {\Delta}^\Phi\right) + \Sigma_{\Xi}{\Delta}^{\Lambda}\right)+ g(1). \nonumber
\eeqn
Therefore, it follows form (\ref{eq:g1}) that the expected cost increase induced by the $\alpha^*=1$ strategy is
\beqn
\mathds{E}\left[J^1\left(\Ac_k\right)\right]-\mathds{E}\left[J^*\left(\Fc_k\right)\right] &&\nonumber \\
&&\nqquad\nqquad \nqquad\nqquad \nqquad =\underbrace{X_k^{\sf T} \Fm^{\sf T}\Gm_{\mathcal{F}_k}^{-1} \left( {\Delta}^{\Gamma}\left({\bf I}- 2\bar{\Vs}\right) - {\Psi} \right)\Gm_{\mathcal{F}_k}^{-1}\Fm X_k}
%%
%_{\succ 0} 
_{> 0} 
+\!\!\! \underbrace{X_k^{\sf T} \Fm^{\sf T} \Gm_{\mathcal{F}_k}^{-1}\bar{\Vs} \Fm X_k}_{\mathds{E}\left[J^0\left(\Ac_k\right)\right]-\mathds{E}\left[J^*\left(\Fc_k\right)\right]}, \nonumber 
\eeqn
where the first term is strictly positive following the assumption that the $\alpha^*=1$ attack construction is optimal. Therefore, $\alpha*=1$ is strictly greater than the $\alpha^*\rightarrow0$ attack strategy only when ${\Delta}^{\Gamma}\left({\bf I}- 2\bar{\Vs}\right) \succ {\Psi}$, which is the condition needed for the $\alpha^*=1$ construction to be optimal.
\section{Non-Stationary Random attacks}
Since the plant given in (\ref{eq1}) is Markovian, it seems reasonable to assume that the attacker should be able to exploit the memory of the system in the construction of the attack. In that sense, the IID attack construction does not provide sufficient flexibility to incorporate the time dependency between consecutive packet losses. Motivated by this insight, in the following we investigate the extension of random attacks to non-IID settings. Specifically, we consider the case in which the statistics of the attack are non-stationary. The resulting non-stationary attack construction extends the IID attack construction to an attack that corrupts a system with independent actuator channels. As in the IID case, the aim of the attacker is to increase the cost function while remaining in the safe operation region by adjusting the value of $\Mm^\alpha$. Additionally, the attack construction is no longer restricted to a constant $\Mm^\alpha$, i.e. $\Mm_k^\alpha\eqdef\mathds{E}[\Vm^\alpha_k]$ for $k\in\mathds{N}$. The derivation of the non-stationary attack construction is equivalent to the IID attack construction up to (\ref{eq:basic}). The reason for the necessity of a different derivation stems from the fact that $\bar{{\Vs}}^\alpha\neq\alpha{\bf I}$ since $\Mm^\alpha_{k}\neq\Mm^\alpha_{k+1}$ for $k\in\mathds{N}$.

We first consider the non-stationary attack construction for the UDP-like protocol. Notice that for the non-stationary construction maximising (\ref{eq:basic}) is equivalent to is equivalent to maximising the function
%Due to the lack of dependence on $\bar{\Vs}^\alpha$ maximising (\ref{eq:basic}) is equivalent to
%
%
\beqn
&&\!\!  f\left(\bar{{\Vs}}^\alpha\right) =\! \Upsilon_{\mathcal{G}_k}^{*{\sf T}}\bar{\Vs}^\alpha\!\! \left( {\Delta}^\Gamma \bar{\Vs}^\alpha +{\Psi} +\left({\bf I}\odot {\Delta}^{\Gamma}\right)\left({\bf I}- \bar{\Vs}^{\alpha}\right)\vphantom{\bar{\Vs}^\alpha}-2\Gm_{\mathcal{G}_k}\right)\Upsilon_{\mathcal{G}_k}^*,\nonumber\\ 
&&\!\! =\mbox{tr}\left(\!\! \bar{\Vs}^\alpha\!\! \left( \left({\Delta}^\Gamma\!\!-\! \left({\bf I} \odot{\Delta}^\Gamma \right)\right)\!\!\bar{\Vs}^\alpha +\left({\bf I} \odot{\Delta}^\Gamma \right) +\!\! {\Psi}\! - 2 \Gm_{\mathcal{G}_k}\right)\Upsilon_{\mathcal{G}_k}^{*}\Upsilon_{\mathcal{G}_k}^{* {\sf T}} \right), \nonumber 
%&& \mbox{s.t.}\qquad \!\!{\Mm}_k^{\alpha} \in \Cc\left(\Mm,\Lm\right) \qquad \textnormal{for all } k=1,2,\ldots,N.
%
%&=& \mbox{tr}\left( \left[ \bar{\Vs}^\alpha\left({\Delta}^\Gamma- \left({\bf I} \odot{\Delta}^\Gamma \right)\right)\bar{\Vs}^\alpha - \bar{\Vs}^\alpha\left(\left({\bf I} \odot{\Delta}^\Gamma \right) + {\Psi} + 2 \bar{{\Vs}}\left({\Delta}^\Gamma -\left({\bf I} \odot{\Delta}^\Gamma \right)\right)\ \right)\right]\Upsilon_{\mathcal{G}_k}^{*}\Upsilon_{\mathcal{G}_k}^{* {\sf T}} \right)
\eeqn
where the maximisation is subject to ${\Mm}_k^{\alpha} \in \Cc\left(\Mm,\Lm\right)$ for $k\in\mathds{N}$.
Letting ${\Delta}^H={\Delta}^\Gamma -\left({\bf I} \odot{\Delta}^\Gamma \right)$, and substituting $\Gm_{\mathcal{G}_k}$ allows the optimal attack strategy for the UDP-like protocol to be posed as a quadratic optimisation problem (QP) given by
\beqn
&& \max_{\bar{{\Vs}}^\alpha}\quad \mbox{tr}\left( \left[ \bar{\Vs}^\alpha{\Delta}^H \bar{\Vs}^\alpha - 
\bar{\Vs}^\alpha\left(\left({\bf I} \odot{\Delta}^\Gamma \right) + {\Psi} + 2 \bar{{\Vs}}{\Delta}^H \right)\right]\Upsilon_{\mathcal{G}_k}^{*}\Upsilon_{\mathcal{G}_k}^{* {\sf T}} \right), \nonumber \\
&& \mbox{s.t.}\qquad \!\!{\Mm}_k^{\alpha} \in \Cc\left(\Mm,\Lm\right) \qquad \textnormal{for } k\in\mathds{N}.
  \eeqn
Note that the set of IID attack strategies is a subset of the strategies generated with this formulation. Therefore, if IID is indeed the optimal attack then the proposed non-stationary attack construction coincides with the strategy presented in the previous section. If however, the cost induced by the non-stationary attack is greater than that induced by $\alpha{\bf I}$, then it follows that memory in the attack yields larger cost increases while satisfying the same detection constraints. Performing the same analysis for the TCP-like system results in an analogous QP formulation given by
\beqn
&& \max_{\bar{{\Vs}}^\alpha}\quad \mbox{tr}\left( \left[ \bar{\Vs}^\alpha{\Delta}^\Gamma \bar{\Vs}^\alpha - 
\bar{\Vs}^\alpha\left(2 \bar{{\Vs}}{\Delta}^\Gamma + {\Psi} \right)\right]\Upsilon_{\mathcal{F}_k}^{*}\Upsilon_{\mathcal{F}_k}^{* {\sf T}} \right), \nonumber \\
&&  \mbox{s.t.}\qquad\!\! {\Mm}_k^{\alpha} \in \Cc\left(\Mm,\Lm\right) \qquad \textnormal{for } k\in\mathds{N}.
\eeqn
%
%\sum_{i=k}^{Nm}
%
In the TCP-like scenario the terms ${\Delta}^\Gamma$, ${\Psi}$, and $\Upsilon_{\mathcal{F}_k}^{*}\Upsilon_{\mathcal{F}_k}^{* {\sf T}}$ are positive semidefinite. Both UDP-like and TCP-like QP formulations can be modified to include IID attacks on non-scalar systems provided that the additional constraint $\Mm^\alpha_k=\Mm^\alpha_{k+1}$ for $k\in\mathds{N}$ is included.
\section{Numerical Results}
\begin{figure}[!t]						%%TCP Scalar!
	\captionsetup{justification=centering,margin=2cm,width=\linewidth}
		\includegraphics[width=0.9\linewidth]{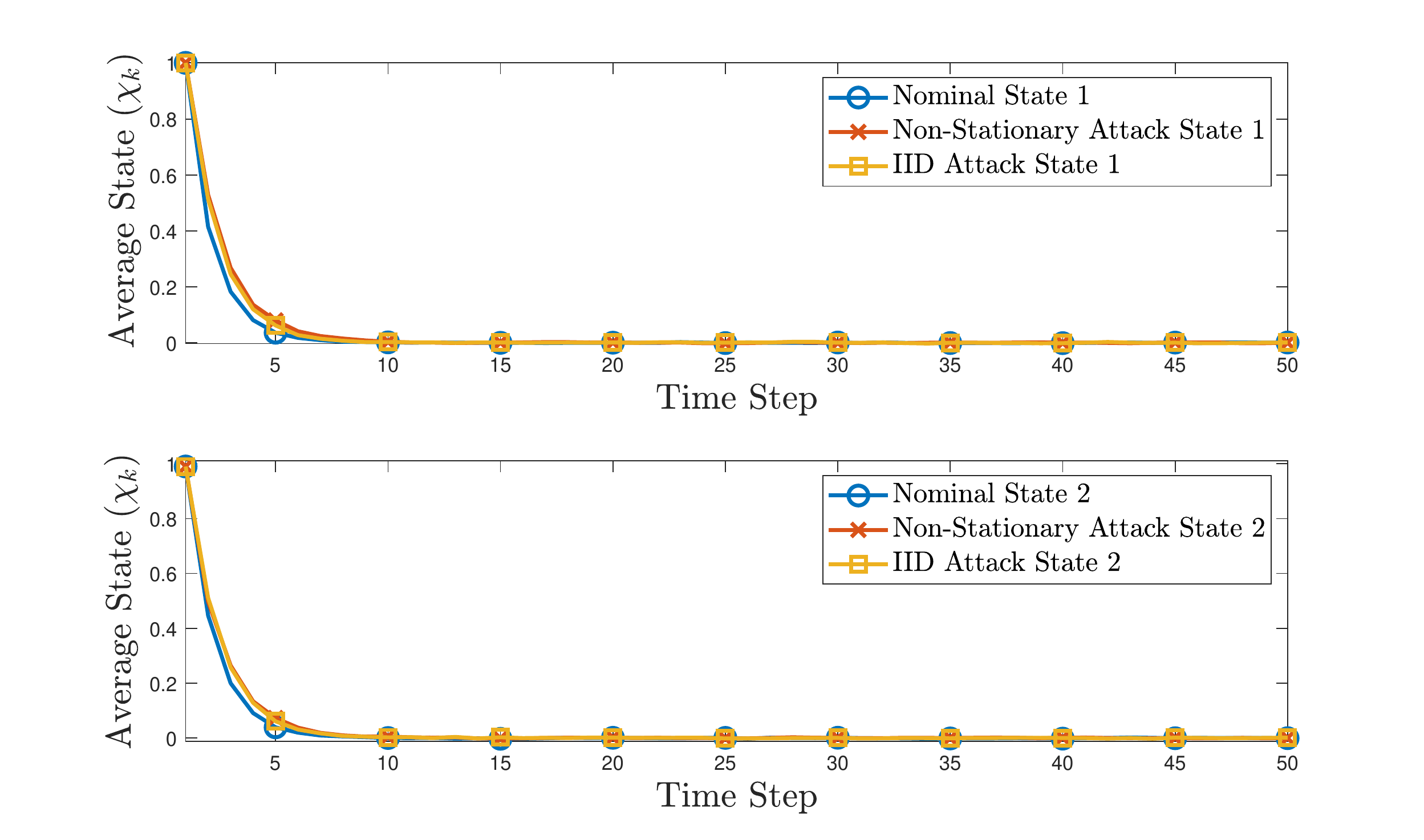}
	\caption{System with TCP-like protocol with $ {\bf A} =\usebox{\smlA}$, $\Mm=0.7$, and $\epsilon=0.1$.}\label{fig3}
\end{figure}

\begin{figure}[!t] 					%% UDP Scalar!
	\captionsetup{justification=centering,margin=2cm,width=\linewidth}
	\includegraphics[width=0.9\linewidth]{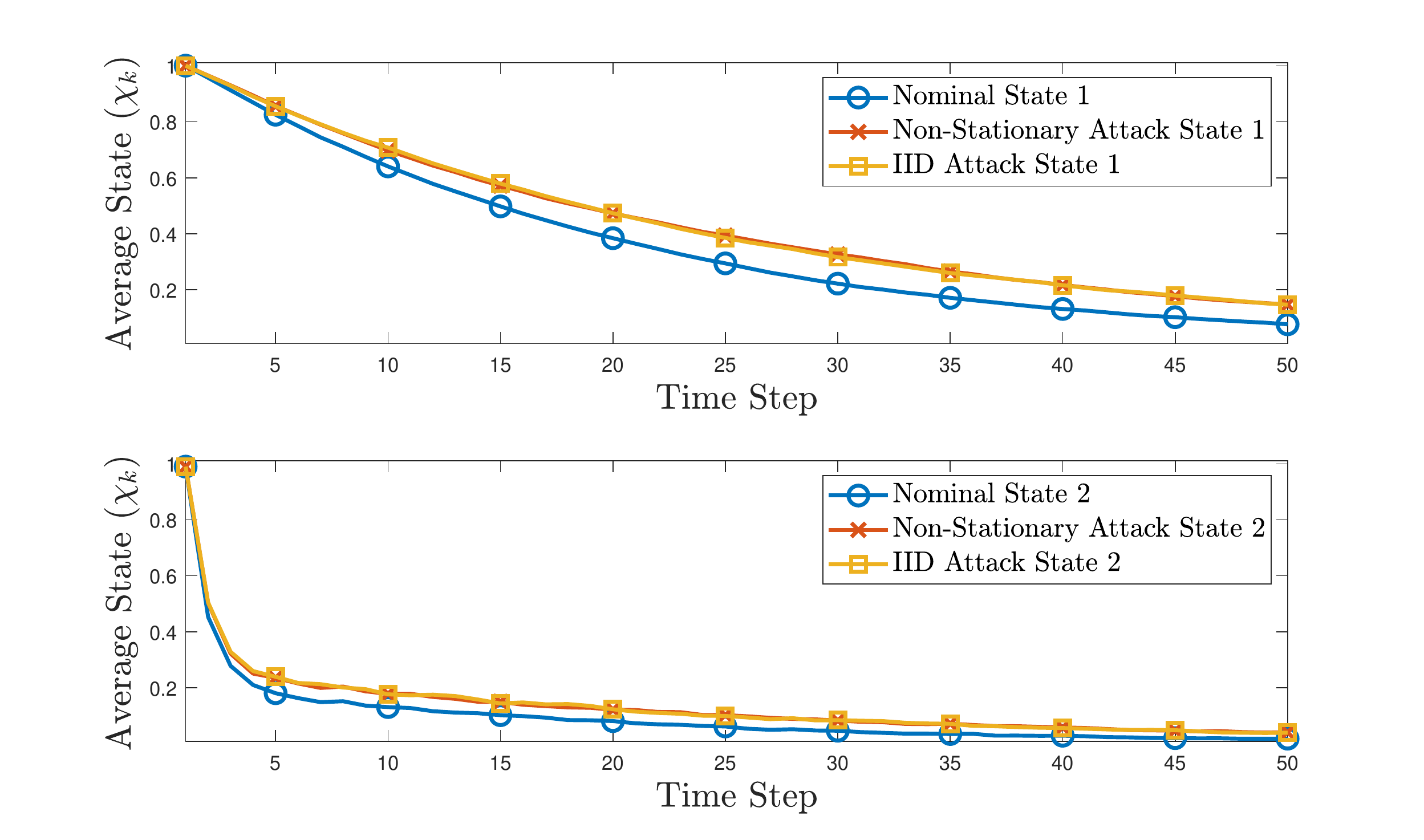}
	\caption{System with UDP-like protocol with ${\bf A} =\usebox{\smlA}$, $\Mm=0.7$, and $\epsilon=0.1$.}\label{fig3a}
\end{figure}

\begin{figure}[!t] 						%% TCP !
	\captionsetup{justification=centering,margin=2cm,width=\linewidth}
	\includegraphics[width=0.9\linewidth]{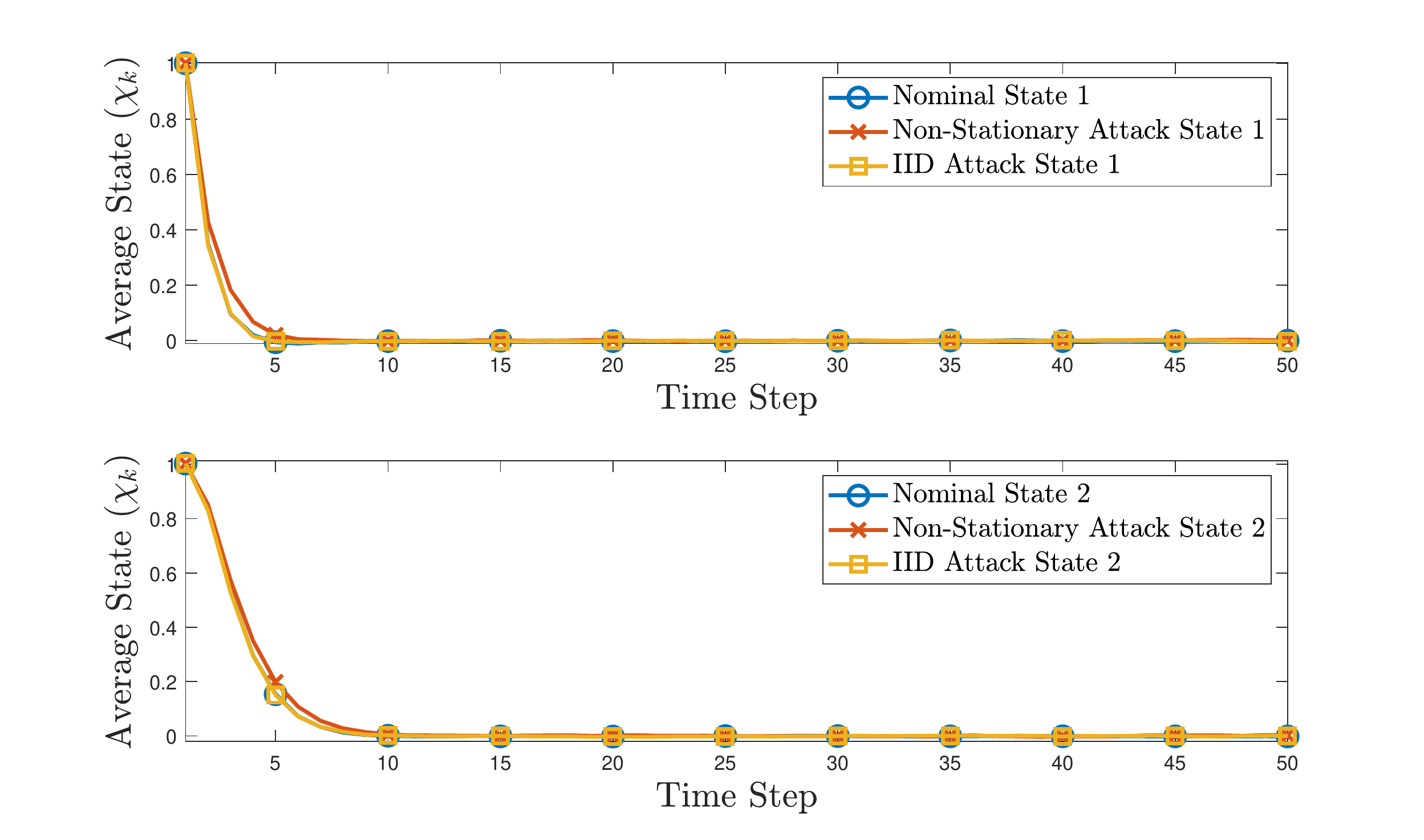}
	\caption{System with TCP-like protocol with ${\bf A} =\usebox{\smlA}$, $\Mm=\usebox{\smlV}$, and $\Lm=0.1 {\bf I}$.}\label{fig4}
\end{figure}

\begin{figure}[!t] 					%% UDP !
	\captionsetup{justification=centering,margin=2cm,width=\linewidth}			
	\includegraphics[width=0.9\linewidth]{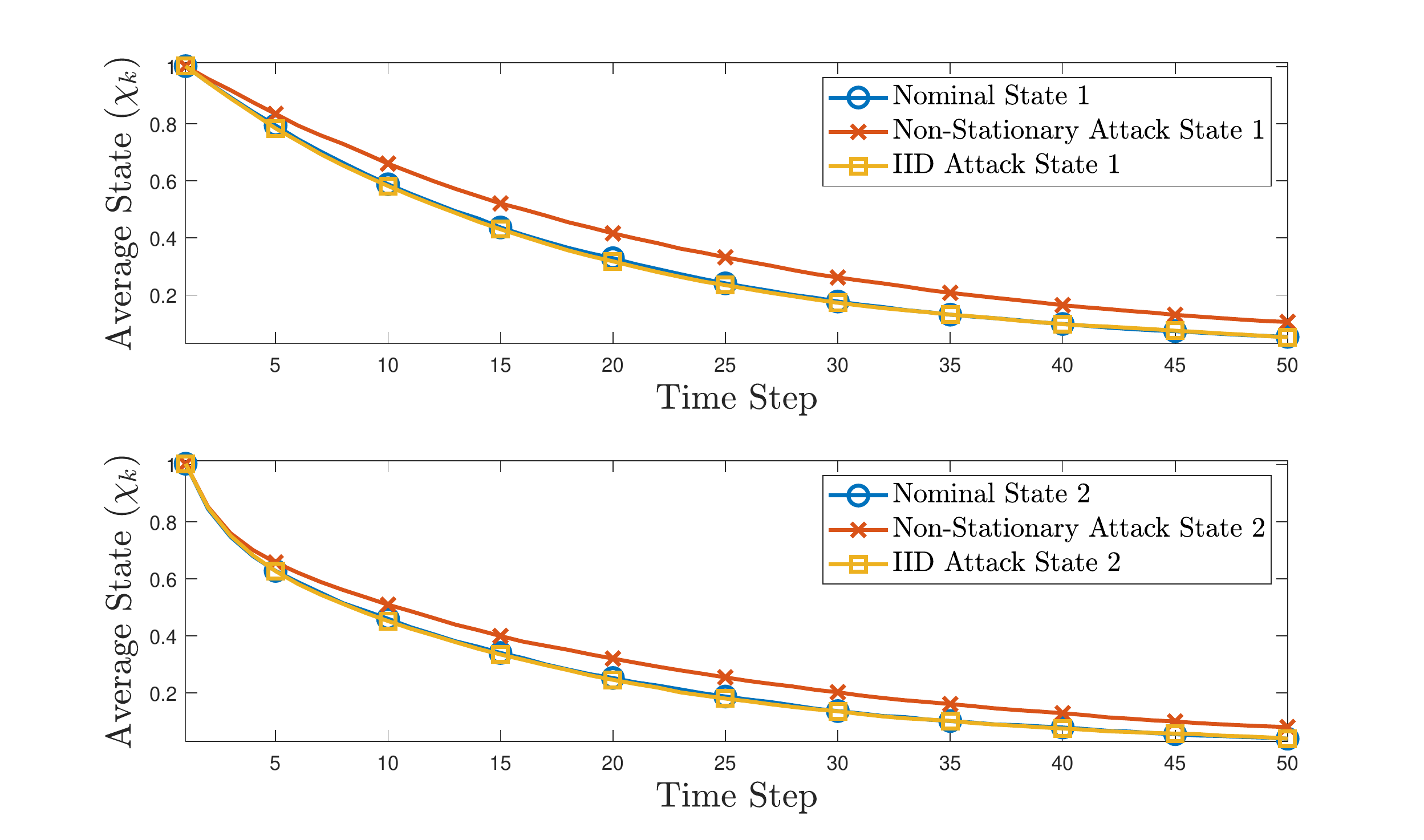}
	\caption{System with UDP-like protocol with ${\bf A} =\usebox{\smlA}$, $\Mm=\usebox{\smlV}$, and $\Lm=0.1{\bf I}.$}\label{fig4a}
\end{figure}

The comparison between the IID attack and the non-stationary attack is conducted over two communication channels for the same system. To that end, we use the same test system as in~\cite{MPC}. The first simulation is performed over a scalar communication channel with $\Mm=0.7$ by averaging $1000$ realisations of the state trajectories and is shown in Fig. \ref{fig3} and \ref{fig3a}.  Additionally, when under attack the UDP-like system trajectory depicted in Fig. \ref{fig3a} displays a larger change from the nominal state trajectory when compared with TCP-like trajectory depicted in  Fig. \ref{fig3}. For the second channel model, i.e. $\Mm=\usebox{\smlV}$, the non-stationary attack results in a larger increase from the nominal state trajectory when compared to the IID attack as shown in Fig. \ref{fig4} and  Fig. \ref{fig4a}, which are obtained by averaging $1000$ realisations of the state trajectories. As in the previous case, the results suggest that the system operating with a UDP-like protocol is more vulnerable to attacks than a system operating with a TCP-like protocol. 
The nominal terminal cost of the system with no attack present is $7.671$ for the UDP-like protocol shown in Fig. \ref{fig4a}. Interestingly, the terminal cost induced by the non-stationary attack is $18.217$ while the terminal cost induced by the IID attack is $13.26$. This suggests that UDP-like protocols are more vulnerable to non-stationary attacks than to IID attacks. However, the difference in the induced cost for the scalar channel shown in Fig. \ref{fig3a} is not as significant as that for the multiple input channel. Surprisingly, for the TCP-like case shown in Fig. \ref{fig4} the performance of the IID attack outperforms the non-stationary attack. Specifically, the terminal cost induced by the IID attack is $8.689$ whereas the terminal cost induced by the non-stationary attack is $8.525$. This small difference is due to numerical error in the simulations, indeed when the accuracy is increased the non-stationary cost converges to the IID attack cost. These results seem to suggest that there is no significant advantage in implementing non-stationary attacks in TCP-like systems. 


%\alert{
%Terminal cost averaged over 1000 realisations for matrix $\Vm$, (UDP-like)
%Non-stationary $18.217$.
%IID Attack, $13.2599$
%Nominal $7.6711$}
%
%\alert{
%	Terminal cost averaged over 1000 realisations for scalar $\Vm$, (UDP-like)
%	Non-stationary $8.862$.
%	IID Attack, $8.695$
%	Nominal $7.736$}
%
%\alert{
%Terminal cost averaged over 1000 realisations for matrix $\Vm$, (TCP-like)
%Non-stationary $8.5251$.
%IID Attack, $8.6890$
%Nominal $7.3483$
%}
%
%\alert{
%	Terminal cost averaged over 1000 realisations for scalar $\Vm$, (TCP-like)
%	Non-stationary $7.3697$.
%	IID Attack, $7.3687$
%	Nominal $7.3473$
%}
%	
\section{Conclusion and Comparison}

We have characterised the optimal IID attack construction for UDP-like and TCP-like systems with lossy actuation channels. The attacks are envisioned as DoS attacks over the actuation communication channel which results in packet losses being induced by the attacker. Under the assumption that the operator monitors the state of the channel with the average packet loss as the decision statistic, we have shown that the optimal attack strategy does not always increase the number of packet losses. In fact, we have characterised the effect of the system parameters over the solution structure and shown that  three different scenarios emerge for which the attack strategy is different. 
Interestingly, under both protocols the attacker only needs to know ${\Delta}^{\Gamma} $, ${\Psi}$, and $\Mm$ to decide the optimal strategy, unless the system operates with a UDP-like protocol and the function is concave, in which case all system parameters must be known. For all cases, the cost increase of the optimal IID construction has been characterised and analysed. We have also shown that the IID attack construction is not optimal by proposing an achievability scheme that constructs attacks with non-stationary statistics. It is shown numerically that the proposed non-stationary attack outperforms the IID attack in most settings although at the expense of increased computational complexity.

\bibliographystyle{ifacconf}
\bibliography{ifac_ref}

\end{document}

%% file: macros.tex
\long\def\comment#1{}

\newfont{\bbb}{msbm10 scaled 700}

\newfont{\bb}{msbm10 scaled 1100}

\newcommand{\Am}{{\bf A}}
\newcommand{\Bm}{{\bf B}}

\newcommand{\Em}{{\bf E}}
\newcommand{\Fm}{{\bf F}}
\newcommand{\Gm}{{\bf G}}

\newcommand{\Lm}{{\bf L}}
\newcommand{\Mm}{{\bf M}}

\newcommand{\Qm}{{\bf Q}}

\newcommand{\Vm}{{\bf V}}

\newcommand{\Ac}{{\cal A}}
\newcommand{\Bc}{{\cal B}}
\newcommand{\Cc}{{\cal C}}

\newcommand{\Fc}{{\cal F}}
\newcommand{\Gc}{{\cal G}}

\newcommand{\Ic}{{\cal I}}

\newcommand{\eqdef}{\stackrel{\Delta}{=}}

\renewcommand{\vec}{{\rm vec}}